\documentclass[aps,pre,twocolumn,superscriptaddress]{revtex4-1}

\usepackage{graphicx,upgreek,amsmath,amsfonts,textcomp, wasysym}
\usepackage[french,english]{babel}
\usepackage[utf8x]{inputenc}
\usepackage[T1]{fontenc}

\usepackage{subfigure}
\usepackage{t1enc,latexsym,amssymb,empheq}
\usepackage{bm}
\usepackage{times}
\usepackage{setspace}
\usepackage{amsthm}
\usepackage{url}

\newcommand{\figref}[1]{Fig.~\ref{#1}}
\renewcommand{\eqref}[1]{Eq.~(\ref{#1})}
\newcommand{\secref}[1]{Sec.~\ref{#1}}

\usepackage[usenames]{color}

\usepackage[usenames,dvipsnames]{xcolor}

\usepackage{color}
\definecolor{darkblue}{rgb}{0,0,0.6}
\definecolor{darkred}{rgb}{0.6,0,0}

\usepackage{hyperref}
\hypersetup{
bookmarksopen=true,
pdftitle="Statistics of roughness for fluctuating interfaces: A survey of different scaling analyses",
pdfauthor="JGuyonnet-EAgoritsas-PParuch-SBustingorry", 
pdfsubject="", 
pdfstartview={FitH},		
pdfmenubar=true,			
pdfhighlight=/O,			
colorlinks=true,			
urlcolor=darkblue,
citecolor=darkblue,		
linkcolor=MidnightBlue,	
}
\begin{document}


\title{Statistics of roughness for fluctuating interfaces: A survey of different scaling analyses}

\author{J. Guyonnet}
\affiliation{DQMP, University of Geneva, 24 Quai Ernest Ansermet, CH--1211 Geneva 4, Switzerland}
\author{E. Agoritsas}
\affiliation{Institute of Physics, EPFL, CH--1015 Lausanne, Switzerland}
\author{P. Paruch}
\affiliation{DQMP, University of Geneva, 24 Quai Ernest Ansermet, CH--1211 Geneva 4, Switzerland}
\author{S. Bustingorry}
\affiliation{Instituto de Nanociencia y Nanotecnología, CNEA--CONICET, Centro At\'omico Bariloche, (R8402AGP) S. C. de Bariloche, R\'{\i}o Negro, Argentina}

\date{\today}


\begin{abstract}

Ferroic domain walls are known to display the characteristic scaling properties of self-affine rough interfaces.
Different methods have been used to extract roughness information in ferroelectric and ferromagnetic materials. 
Here, we review these different approaches, comparing roughness scaling analysis based on displacement autocorrelation functions in real space, both locally and globally, to reciprocal space methods.
This allows us to address important practical issues such as the necessity of a sufficient statistical averaging.
As an ideal, artifact-free reference case and particularly targeting finite-size systems, we consider two cases of numerically simulated interfaces, one in equilibrium with a disordered energy landscape and one corresponding to the critical depinning state when the external applied driving force equals the depinning force.
%
We find that the use of the reciprocal space methods based on the structure factor allows the most robust extraction of the roughness exponent when enough statistics is available, while real space analysis based on the roughness function allows the most efficient exploitation of a dataset containing only a limited number of interfaces of variable length.
This information is thus important for properly quantifying roughness exponents in ferroic materials.

\end{abstract}


\maketitle

\section{Introduction}
\label{sec_intro}

Ferroic materials are characterized by a spontaneous order parameter that can be reversibly switched between at least two energetically-equivalent ground states by an appropriate conjugated field. For example, in ferroelectrics and ferromagnets these order parameters are the polarization and the magnetization, respectively, switchable by applying an electric or magnetic field. Regions of homogeneous order parameter state in the sample are called domains, separated by nanoscale boundaries known as domain walls. The ability to controllably engineer ferroic domains in increasingly miniaturised devices has played a significant role in the integration of these materials into the electronics industry~\cite{polla_97_mems,kumar_apl_04_SAW, scott_sci_89_memories,waser_natmat_04_memories, Parkin2008, Hayashi2008, Allwood2005}. At the most fundamental level, such engineering is built on the understanding and control of the static and dynamical behavior of the domain walls, which determine the switching, growth, stability, and shape of ferroic domains \cite{lemerle_prl_98_FMDW_creep, paruch_jap_06_dynamics_FE, paruch_dw_review_07, metaxas_depinning_thermal_rounding, Metaxas2010}. 

One extremely useful theoretical approach to study domain walls in ferroic materials is to model them as fluctuating elastic manifolds subject to the spatial inhomogeneities of an underlying disordered potential~\cite{giamarchi_domainwall_review}.
A remarkable feature of this reductionist picture is that, because the underlying microscopic details of the system are only considered through a few effective parameters, it can be applied to systems as diverse as surface growth phenomena~\cite{barabasi_surface_growth_95}, fracture surfaces~\cite{mandelbrot_nature_84_cracks_metal}, burning~\cite{myllys_prl_00_burning_fronts} and wetting~\cite{rubio_prl_89_wetting_fronts} fronts, edges of bacterial colonies~\cite{bonachela_jstatphys_11_bacterias_DES}, cell migration \cite{Chepizhko_pnas_16_cell_front}, cell membranes~\cite{speck_pre_12_cell_membrane}, as well as ferroic domain walls~\cite{lemerle_prl_98_FMDW_creep,paruch_13_FE_DW_review,Ferre_cras_13_DW}.
In this approach, the complex static and dynamical properties of the interface emerge from a seemingly simple competition between elasticity and disorder pinning.  In particular, such \emph{disordered elastic systems} present a rough morphology with characteristic self-affine scaling properties, which depend on the dimensionality of the system, the range of the elastic interactions, and the nature of the disorder~\cite{agoritsas_physb_12_DES}.
The quantitative characterization of this roughness, including the value of the associated scaling roughness exponent $\zeta$, can rely on several methods either in real or in reciprocal space, and choosing among them is thus a key issue of roughness analyses.

Experimental roughness studies in ferroic materials~\cite{lemerle_prl_98_FMDW_creep, paruch_prl_05_dw_roughness_FE, Bauer2005, catalan_prl_08_BFO_DW, pertsev_jap_11_ceramics, xiao_apl_13_PVDF_roughness_creep} have generally used real-space analysis of such domain walls, built on images covering a finite number of pixels, typically of the order of a few hundreds, and thus always requiring a detailed assessment of finite-size effects.
More importantly, real-space methods are mainly used to extract the value of the roughness exponent $\zeta$ from the power-law growth of the correlation function of relative displacements, with the corresponding uncertainty resulting from statistical averaging.
However, as shown in a comparative study of analysis methods on numerical --~and thus exactly defined~-- self-affine profiles, the accuracy of $\zeta$ estimation can in fact vary significantly depending on the method used~\cite{schmittbuhl_pre_95_reliability_roughness}.
Furthermore, adequate statistical averaging is an absolutely critical issue, with trustworthy $\zeta$ estimates obtained only when considering at least a few tens of independent mono-affine interfaces~\cite{guyonnet_prl_12_multiscaling}. 
Finally, we note that while static pinned interfaces in equilibrium are very well understood, much less is known both theoretically and experimentally about the configurations of a moving interface, in the creep, depinning, or linear flow regimes, which may be characterized by more complex behavior, such as super-roughening~\cite{rosso_prl_03_depinning,bustingorry_prb_12_depinning}.
In such cases, analyses based on real-space displacement field correlations can lead to erroneous results~\cite{leschhorn_prl_93_superroughening}.
In ferroelectrics for example, unusual roughening has actually been observed in high-velocity driven domain walls, with local curvature acting as a precursor for the ejection of nanodomains~\cite{shur_jetp_91_skyrmions_ferroelectric}, possibly skyrmion-like in character~\cite{dawber_jpcm_06_DW_skyrmions}.  

A quantitative evaluation of the different roughness analysis methods, taking into account not only ideal numerical systems but also the frequent experimental limitations on the size and number of interfaces available for study, and the possibility of more complex roughness scaling, would therefore be evidently very useful in resolving many of these open questions. Such an evaluation would also allow the establishment of a well-defined analysis protocol, which could be applied over all the different ferroic systems under investigation and moreover to the general class of interfaces described as disordered elastic systems~\cite{barabasi_surface_growth_95, Brazovskii2004, giamarchi_domainwall_review, agoritsas_physb_12_DES}.

We thus perform here an evaluation of different analysis methods using two different models of fluctuating interfaces. Though the different methods we consider are essentially complementary, some key features are worth to be noted. First, the roughness exponent obtained as the mean value of individual roughness exponents, \textit{i.e.} each from a single independent interface, converges to the roughness exponent obtained using averaged correlation functions. Second, the distribution of roughness exponents from independent interfaces, an experimentally relevant measure, is appreciably wide and size-dependent. Finally, our results suggest that a set with at least 40 independent interface realizations should be used to obtain representative averaged values for the roughness exponent, something usually overlooked in experimental reports.

The manuscript is organized as follows. In Sec.~\ref{sec_roughness_analysis} with a definition of the different roughness analyses methods, and a discussion of some of the unresolved issues associated with more complex behavior.
We then compare these different analysis methods in Sec.~\ref{sec_numerical} using a numerical one-dimensional model system with an exactly known interface position and no experimental artifacts as a test, and focusing on small system sizes.
We first focus on the ideal case of pinned equilibrated interfaces in a random-bond disorder, characterized by a mono-affine roughness scaling and a known value of $\zeta$. Subsequently, we turn to the more complex case of driven interfaces at the critical depinning transition, which exhibits super-roughening behavior.
Finally, a summary of the results is presented in Sec.~\ref{sec_concl}.

\section{Roughness scaling analysis}
\label{sec_roughness_analysis}

Since the seminal work of Mandelbrot \textit{et al.} revealing the self-affine nature of cracks in metals~\cite{mandelbrot_nature_84_cracks_metal}, a significant number of different methods have been established and used to quantify the roughness of self-affine interfaces, focusing in particular on fracture surfaces~\cite{maloy_prl_92_crack,schmittbuhl_jgeophys_95_crack_scaling}. 
The key quantity to be determined is the roughness exponent $\zeta$, which characterizes the geometrical properties of interfaces through the power-law growth of their transverse fluctuations $w$ with respect to the longitudinal size of the interface $\ell$, \textit{i.e.} ${w \sim \ell^\zeta}$.
In all the proposed methods for the determination of $\zeta$, a complete knowledge of the interface position is assumed, in which case the analysis of the roughness can be carried out via either reciprocal-space or real-space autocorrelation functions.

In this section, we briefly recall the different definitions of the roughness, and discuss some important issues regarding their analysis and comparison.
We restrict ourselves to effective one-dimensional (1D) interfaces, as they are particularly relevant for many experimental ferroic domain walls, but the following definitions can be generalized to higher dimensions.

\subsection{Measuring the roughness exponent $\zeta$}

In a general sense, the roughness of an interface characterizes its geometrical fluctuations~\cite{barabasi_surface_growth_95}.
Here we specify different definitions of correlation functions giving alternative access to the roughness exponent, all relying on the displacement field ${u(z,t)}$ which parameterizes a given configuration of an interface at time $t$ with respect to an arbitrary reference configuration, as illustrated in Fig.~\ref{fig:def-roughness}(a).
A usual assumption in the theoretical framework of disordered elastic systems is that the interface has no overhangs, so that ${u(z,t)}$ is univalued~\cite{agoritsas_physb_12_DES}.

\subsubsection{Global width ${W(L,t)}$}

The geometrical roughness of an interface may first be quantified by the system-size dependence of the perpendicular fluctuations of the interface position around its mean value, more specifically by their variance, a quantity referred to as the \emph{global width}:
\begin{equation}
 W(L,t)
 = \overline{\langle \left[ u(z,t)-\langle u (z,t) \rangle_L \right]^2 \rangle_L}^{1/2},
\label{equ_WL}
\end{equation}
where $L$ is the system size, $z$ is the longitudinal spatial coordinate, ${\langle\cdots\rangle_L}$ denotes the spatial averaging over the entire interface, \textit{e.g.} the mean position ${\langle u(z,t) \rangle_L = L^{-1} \int_0^{L} dz \, u(z,t)}$, and $\overline{\cdots}$ denotes disorder averaging obtained in practice by averaging over multiple interfaces.

For most surface growth phenomena evolving from a flat initial configuration, there exists a longitudinal correlation length $\xi(t)$ that increases with time.
In the absence of characteristic length scales other than $L$ and $\xi$, a power-law behavior in space and time can be expected and ${W(L,t)}$ satisfies the Family-Vicsek scaling ansatz~\cite{FamilyVicsek}
\begin{equation}
 W(L,t)\sim
 \left
 \{
 \begin{array}{ll}
 t^{\zeta/z_d} & \mathrm{for\ }\xi(t) \ll L\\
 L^{\zeta} & \mathrm{for\ } \xi(t)\gg L,
 \end{array}
 \right.
\label{equ_WL_FV}
\end{equation}
where $z_d$ is the dynamical exponent characterizing the growth of the longitudinal correlation length ${\xi(t) \sim t^{1/z_d}}$, and $\zeta$
is referred to as the roughness exponent and characterizes the stationary regime, in which the longitudinal correlation length ${\xi(t)}$ has reached a value close to the system size $L$ and ${W(L,t)}$ becomes time-independent.
For stationary systems such as interfaces in equilibrium, the time dependence may thus be dropped, simplifying the formalism.

In practice, a direct measurement of $\zeta$ through ${W(L,t)}$ requires precise control of the system size~$L$ over several orders of magnitude  and a large number of interfaces for sufficient averaging, which can be very difficult to achieve in experiments.

\subsubsection{Local width ${w(r,t)}$}

\begin{center}
\begin{figure}[!thbp]
\includegraphics[width=\columnwidth]{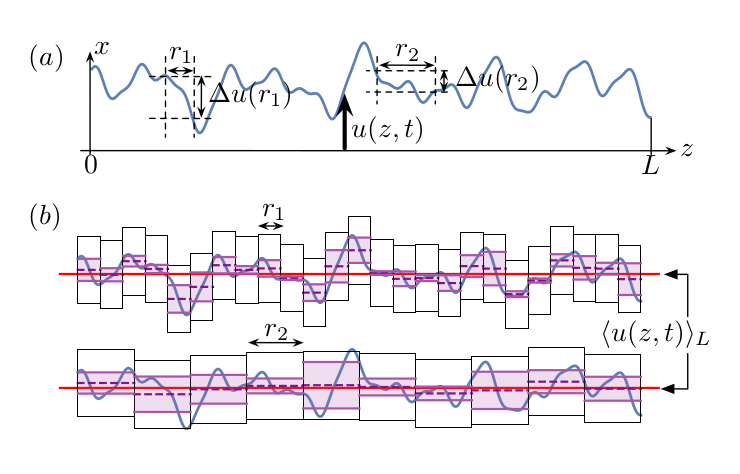}
\caption{
(a)~Profile of a 1D interface, parameterized by the displacement field ${u(z,t)}$. The variance of its relative displacements ${\left\lbrace \Delta u(r,t) \right\rbrace}$ is given by the displacement-displacement correlation function ${B(r)}$.
(b)~Schematic illustration of the local width ${w(r,t)}$ for ${r_1 < r_2}$. In each box, the dotted line indicates the mean position ${\langle u(z,t) \rangle_r}$ of the corresponding segment (which fluctuates with respect to the global mean ${\langle u(z,t) \rangle_L}$), and the dashed area its standard deviation whose average gives ${w(r,t)}$.
}
\label{fig:def-roughness}
\end{figure}
\end{center}

It has been found that in many cases, local quantities measuring the interface fluctuations over a smaller window ${r<L}$ 
also exhibit a power-law behavior as a function of $r$, characterized by the same scaling exponents.
One of these quantities is the \emph{local width}:
\begin{equation}
 w(r,t)
 = \overline{\langle \left[ u(z,t)-\langle u(z,t) \rangle_r \right]^2 \rangle_r}^{1/2},
\label{equ_wr}
\end{equation}
where ${\langle \cdots \rangle_r}$ is a spatial average over windows ${z \in D_r}$ of size $r$, \textit{e.g.}~${\langle u(z,t) \rangle_r = r^{-1} \int_{z \in D_r} dz \, u(z,t)}$.
This gives essentially the average of the width of the portions of the interface in a window of size $r$, and the disorder average is provided here by averaging over the $N_r$ such windows available on a given interface, as illustrated in Fig.~\ref{fig:def-roughness}.
The local width is the natural generalization of its global counterpart, and allows us to probe the geometrical fluctuations as a function of length scale, exploiting more of the information that is provided by the displacement field ${u(z,t)}$.
The global width is then the limiting case ${w(r=L,t)=W(L,t)}$.

Due to scale invariance, it is expected that when there is only one characteristic length scale, the local width grows with the window size as
\begin{equation}
 w(r,t)\sim
 \left
 \{
 \begin{array}{ll}
 t^{\zeta/z_d} & \mathrm{for\ } \xi(t) \ll r\\
 r^{\zeta} & \mathrm{for\ } \xi(t)\gg r.
 \end{array}
 \right.
\label{equ_wr_FV}
\end{equation}
The Family-Vicsek scaling property of the global width does not necessarily imply this scaling behavior for the local width~\cite{lopez_pre_97_anomalous_scaling,lopez_pre_98_anomalous_scaling,ramasco_prl_00_generic_scaling} (see \secref{sec_br_superrough}).

\subsubsection{Displacement-displacement correlation function ${B(r,t)}$}

Another local quantity containing geometrical information on interfaces is the \emph{displacement-displacement correlation function}, sometimes referred to as the height-height correlation function, the height-difference correlation function, or simply the roughness function:
\begin{equation}
 B(r,t)
 = \overline{\langle [u(z+r,t) - u(z,t)]^2 \rangle_L},
 \label{equ_br}
\end{equation}
where ${\Delta u_z(r,t) =u(z+r,t) - u(z,t)}$ is the relative transverse displacement between pairs of sites a distance $r$ apart, as illustrated in Fig.~\ref{fig:def-roughness}, and ${B(r,t)}$ is simply the variance of the probability distribution function (PDF) of relative displacements ${\mathcal{P}(\Delta u(r,t) )}$.

For self-affine interfaces with a single characteristic scale~${\xi(t)}$, we have:
\begin{equation}
 B(r,t)\sim
 \left
 \{
 \begin{array}{ll}
 t^{2 \zeta/z_d} & \mathrm{for\ } \xi(t) \ll r\\
 r^{2 \zeta} & \mathrm{for\ } \xi(t)\gg r.
 \end{array}
 \right.
\label{equ_Br_FV}
\end{equation}
This displacement-displacement correlation function provides a convenient way to experimentally measure the roughness exponent, and has thus been used as a primary analysis tool in ferroic systems~\cite{lemerle_prl_98_FMDW_creep,Jost_pre_98_roughness_B,paruch_prl_05_dw_roughness_FE,catalan_prl_08_BFO_DW,paruch_prb_12_quench}.

Dimensionally, we have ${B(r,t) \sim w(r,t)^2}$, but the physical content of these two quantities is fundamentally different, as emphasized side by side in Fig.~\ref{fig:def-roughness}:
${w(r,t)}$ characterizes the fluctuations around the mean position over a segment of size $r$,
whereas ${B(r,t)}$ measures the correlation between two points separated by a distance $r$.

\subsubsection{Structure factor ${S(q,t)}$}
\label{sec_sq}

An alternative option to real-space correlation functions is to compute them in reciprocal space. A particularly useful quantity is the displacement power spectrum, referred to as the \emph{structure factor}:
\begin{equation}
 S(q,t) = \overline{\tilde{u}(q,t) \tilde{u}(-q,t)},
\label{equ_sq}
\end{equation}
where
\begin{equation}
 \tilde{u}(q,t) = \frac{1}{L} \int dz \, u(z,t) \, e^{-i q z}
\end{equation}
is the Fourier transform of the displacement field $u(z,t)$ defining the interface position. Formally, the structure factor ${S(q)}$ and the displacement-displacement correlation function ${B(r)}$ contain the same geometrical information and are related through
\begin{equation}
 B(r,t)
 = \int \frac{dq}{\pi} \, \left[1-\cos\left(qr\right)\right] \, S(q,t).
\label{equ_br_sq}
\end{equation}
For a self-affine interface, the structure factor scales as
\begin{equation}
 S(q,t)\sim
 \left
 \{
 \begin{array}{ll}
 t^{(1+2 \zeta)/z_d} & \mathrm{for\ } q \ll \xi(t)^{-1}\\
 q^{-(1+2 \zeta)} & \mathrm{for\ } q \gg \xi(t)^{-1}.
 \end{array}
 \right.
\label{equ_Sq_FV}
\end{equation}

From a practical point of view, when sufficient statistics can be obtained (with a high resolution, and/or large systems, and/or many systems to average over), fitting ${S(q)}$ has been shown to be a generally more reliable method to estimate $\zeta$ than the real-space autocorrelation functions \cite{schmittbuhl_pre_95_reliability_roughness},
essentially because different scaling regimes depending on the length scale are clearly separated in reciprocal space, whereas they are mixed in real space functions.
Moreover, as we discuss in \secref{sec_br_superrough}, it can be used to determine roughness exponents ${\zeta>1}$ for super-rough interfaces. More fundamentally, the structure factor is a very important quantity for theoretical developments, and has in particular been shown to be pivotal in the formalism of anomalous scaling~\cite{ramasco_prl_00_generic_scaling}.

\subsection{Specific issues of roughness scaling analyses}

\subsubsection{Time and length scales}

Up to now we have given a general discussion of the different quantities of interest and of their scaling properties for both space and time variables.
Such a roughness analysis permits us to analyze the important out-of-equilibrium and dynamical regimes where the time evolution of the interface is a primary target. However, this is not always the case, as for example when analyzing equilibrium static properties, since the system is not evolving and the time scale is dropped as meaningless. When extracting the roughness exponent in experiments, for instance in ferroelectric or ferromagnetic~\cite{lemerle_prl_98_FMDW_creep} domain walls, such a static situation has been generally assumed. This physically corresponds to assuming that the time-dependent longitudinal correlation length scale has become larger than the actual system size, ${\xi(t) \gg L}$.
In the different situations analyzed in the rest of the present work we shall assume that the time scale is large enough to guarantee this condition.
This is strictly true for the two numerical models analyzed in \secref{sec_numerical} but it has to be carefully considered for systems where time stability is an issue, as for small ferroelectric domains~\cite{blaser_apl_12_CNT_FE,paruch_apl_06_stability}.
Therefore, when ${\xi(t) \gg L}$, the time variable is dropped and the roughness exponent can in principle be extracted alternatively from the following relations
\begin{subequations}
\label{equ_scal_zeta}
\begin{align}
 W(L) & \sim L^\zeta \label{equ_scal_zeta_W},\\
 w(r) & \sim r^\zeta \label{equ_scal_zeta_w},\\
 B(r) & \sim r^{2 \zeta} \label{equ_scal_zeta_B},\\
 S(q) & \sim q^{-(1+ 2 \zeta)}. \label{equ_scal_zeta_S}
\end{align}
\end{subequations}

Both in numerical and experimental approaches there is usually an intrinsic small length cutoff associated with either discretization of the $z$ direction in the numerical approach or with resolution issues (pixel size) of images in experiments.
This sets on the one hand the lower length scale limit $a$ and the corresponding large wave vector ${q=2 \pi /a}$.
On the other hand the large scale cutoff is given by the system size $L$ and its corresponding small wave vector ${q=2 \pi/L}$.

Finally, since in the present work we will be considering discretization of the $z$ direction, as is usually the case both in numerical and experimental situations, it is convenient to use a discreteness correction to the wave vector when dealing with the Fourier modes. Consider for instance the discretization as ${z=j \Delta z}$, with ${j=0,1,2,...,L-1}$. Without loss of generality we take ${\Delta z=1}$. Then the interface profile becomes $u_j$ and its Fourier transform is ${\tilde u_n=L^{-1} \sum_{j=0}^{L-1} u_j e^{-i q_n j}}$, with ${q_n=2 \pi n/L}$.
The large wave vector correction due to the discreteness of $z$ is achieved through the discretization of the Laplacian, ${\nabla^2 u(z) \to u_{j+1} - 2 u_j + u_{j-1}}$, which after Fourier transformation becomes $-q^2 \tilde u(q) \to \tilde u_n e^{-i q_n} - 2 \tilde u_n + \tilde u_n e^{i q_n}=-4 \sin^2(q_n/2) \tilde u_n$.
Therefore ${\tilde q_n = 2 \sin(q_n/2)}$ can be identified as the discretization-corrected wave vector properly controlling small length scale behavior, as it has been explicitly used~\cite{kolton_prl_06_DWdepinning,Rosso_prb_07_numericalFRG, Ferrero2013}.

\subsubsection{Scaling of ${B(r)}$ for super-rough interfaces}
\label{sec_br_superrough}

Following earlier reports of roughness scaling analyses~\cite{lopez_pre_98_anomalous_scaling,ramasco_prl_00_generic_scaling,Torres_epjb_13_anomalous_scaling}, a generalized formalism distinguishes the global, local, and reciprocal scaling behaviors of the self-affine interface, with ${W(L)}$, ${w(r)}$ and ${B(r)}$, and ${S(q)}$ scaling according to their corresponding global, local, and reciprocal scaling exponents.
In this picture, standard Family-Vicsek scaling is given by all three exponents being equal. Any other case falls in one of three categories of anomalous scaling (cf. Ref.~\cite{ramasco_prl_00_generic_scaling} for details). In particular, super-rough interfaces are globally characterized by ${\zeta>1}$, as found using ${W(L)}$ and ${S(q)}$, and locally (where Family-Vicsek relations are no longer valid) by ${\zeta_{\mathrm{loc}}=1}$ when using ${w(r)}$ or ${B(r)}$.

Phenomenologically, ${\zeta>1}$ corresponds to the seemingly unphysical case where the transverse fluctuations become unbounded at very large length scales. In such a case, a crossover to a bounded regime may therefore be expected.
Numerically, 1D driven interfaces at the depinning threshold were shown to possess a roughness exponent ${\zeta^{1\mathrm{D}}_{\mathrm{dep,harm}}=1.25}$ when only short-range harmonic contributions to the elastic energy were considered~\cite{Ferrero_pre_13_numerical_exponents}, and ${\zeta^{1\mathrm{D}}_{\mathrm{dep,anharm}}=0.635}$ when an anharmonic correction was taken into account~\cite{rosso_prl_03_depinning}.
Thus, the depinning phase of 1D interfaces should exhibit a crossover from a super-rough regime at small enough length scales to a bounded regime with ${\zeta<1}$.
Experimentally, driven magnetic domain walls were recently shown to exhibit a roughening behavior consistent with this interpretation~\cite{bustingorry_prb_12_depinning, Grassi2018}.
1D static interfaces are also predicted to exhibit such a crossover at small length scales, at least in a `low-temperature' regime~\cite{agoritsas_2010_PhysRevB_82_184207,agoritsas_2012_FHHtri-numerics}.

To understand the discrepancy between the global and local roughness scaling behavior for super-rough interfaces, the analytical expression of ${B(r)}$ can be considered. The usual derivation starts from the relation between ${B(r)}$ and ${S(q)}$ given in~\eqref{equ_br_sq}.
Assuming a long-time Family-Vicsek scaling for ${S(q)}$ (\eqref{equ_Sq_FV}), the scaling behavior of ${B(r)}$ is then given in all generality by
\begin{equation}
 B(r)
 \sim \int_{2\pi/L}^{2 \pi/a} \frac{dq}{\pi} \left[1-\cos\left(qr\right)\right]q^{-(1+2\zeta)},
\label{equ_br_general}
\end{equation}
When ${0 < \zeta \leq 1}$ and taking the limits ${a \to 0}$ and ${L \to \infty}$ the integral converges and the Family-Vicsek scaling relation is recovered, with a single $\zeta$ value describing both the local and global correlation functions. In the case ${\zeta > 1}$, taking ${a \to 0}$ and for large but finite values of $L$, the roughness function ${B(r)}$ obeys the general scaling behavior for `super-rough' interfaces
\begin{equation}
 B_{\mathrm{SR}}(r)
 \approx r^2\left[ -A_1r^{2(\zeta-1)} + A_2L^{2(\zeta-1)} \right],
\label{equ_br_superrough}
\end{equation}
where $\zeta$ is the global roughness exponent, $A_1$ and $A_2$ are positive constants, and ${C(L) = A_2L^{2(\zeta-1)}}$ is an $L$-dependent constant.
The presence of the $r^2$ prefactor indicates that, when taking the limit ${L\to\infty}$ first and then the large $r$ limit, the local roughness exponent saturates to ${\zeta_{\mathrm{loc}} = 1}$.
We note that this expression is slightly more general than the one reported in Ref.~\cite{lopez_pre_98_anomalous_scaling}, which holds only for large values of $L$.
Such a super-rough behavior will be illustrated in Sec.~\ref{sec_num_dep} on numerical simulations.

\subsubsection{The necessity of statistical averaging}
\label{sec_zeta_avg}

As the scaling relations in Eqs.~(\ref{equ_scal_zeta}) only hold with the appropriate statistical averaging, a crucial step in roughness analysis of experimental interfaces is to assess the minimal number of independent configurations necessary to achieve a meaningful estimation of the roughness exponent $\zeta$. Furthermore, one may ask how representative of the actual roughness exponent is the value obtained from a single measurement. In both cases, we can expect the answer to be both size and method-dependent.

For numerical simulations where a large number of independent realizations can be available, a meaningful estimation of $\zeta$ can readily be obtained by computing the desired correlation functions averaged over the number of realizations, and subsequently fitting a single $\zeta$ value from the power-law behavior. In contrast, for experimentally imaged interfaces the amount of different realizations typically ranges in the order of a few tens and may in addition suffer from differences in size and resolution. Therefore, a common practice is to compute the appropriate function for each interface and extract an average value using
\begin{equation}
\overline{\zeta} = \frac{1}{N} \sum_{i=1}^{N} \zeta_i,
\end{equation}
where $N$ is the number of independent measurements and $\zeta_i$ is the roughness exponent of a single interface ${u_i(z)}$ as obtained for example using
\begin{equation}
B_i(r) = \langle \left[u_i(z+r)-u_i(z)\right]^2\rangle \sim r^{2 \zeta_i}.
\end{equation}
It is important to note that $\overline{\zeta}$ is not necessarily equivalent to $\zeta$ unless the underlying distribution for the roughness exponent happens to be symmetric. Therefore, the skewness of the $\zeta_i$ histogram is indicative of the validity of this method and the accuracy of $\overline{\zeta}$~\cite{guyonnet_prl_12_multiscaling}.

We want to emphasize that the exponent $\zeta$ obtained first by averaging the ${B_i(r)}$ and secondly fitting it with a power law is the actual `physical' roughness exponent, in the sense that it corresponds to the quantity defined and computed theoretically.
Nevertheless, considering the histogram of the $\zeta_i$ fitted on individual interfaces is particularly relevant in experimental systems, in order to assess the composition and the quality of a given batch of measurements.
For example, a two-peak histogram would suggest that two groups of measurements should be distinguished in the batch;
and a one-peak histogram with a small variance would indicate that individual measurements have already enough statistics or are sufficiently large to reproduce the predicted overall $\zeta$
of finite size and with a limited statistics.
Following Ref.~\cite{guyonnet_prl_12_multiscaling}, we will show in Sec.~\ref{sec_num_equ} that the value of $\bar{\zeta}$ can actually coincide with $\zeta$, suggesting that both values can be computed in experiments, and their comparison used as a necessary but not sufficient criterion of the validity of a given batch.

\section{Numerical simulations}
\label{sec_numerical}

In this section we study the effects of system size, statistical averaging, and roughness analysis methods to estimate the roughness exponent of numerically simulated 1D interfaces, either in equilibrium or in a critical depinning state.
In both cases, we start by showing that the interfaces are characterized by Gaussian displacements PDF, and are therefore mono-affine. We then show that equilibrium simulations are well described by Family-Vicsek scaling with ${\zeta = 2/3}$, a well-known result~\cite{huse_henley_fisher_1985_PhysRevLett55_2924}.
In contrast, critical depinning configurations are characterized by ${\zeta = 1.26}$, consistently obtained through the global width, structure factor, and appropriate displacements autocorrelation function for super-rough interfaces.
Finally, for the equilibrium case, we show that the average roughness exponent $\overline{\zeta}$ converges towards $\zeta$ for a number of configurations close to 50, with faster convergence for the structure factor ${S(q)}$ than for the displacement-displacement correlation method~${B(r)}$.

\subsection{Interfaces in equilibrium}
\label{sec_num_equ}


Interfaces in equilibrium can be generated by allowing an interface living in a disordered energy landscape to relax to its minimum energy configuration. 1D equilibrated interfaces in weak collective random-bond disorder were simulated from a directed polymer model~\cite{mezard_jdpi_92_manifolds} on a discretized square lattice.
The position of the polymer is given by ${u(z)}$, with $u$ and $z$ taking discrete values, mimicking a fluctuating interface. The solid-on-solid restriction ${|u(z+1) - u(z)| = \pm 1}$ provides the effective short-range elasticity to the polymer model. An uncorrelated Gaussian random potential distributed on each lattice site, ${V(u,z)}$, is used to model a disordered energy landscape. Disordered potential correlations are given by ${\overline{V(u,z)V(u',z')} = D \, \delta_{u,u'}\delta_{z,z'}}$, where $D$ is the intensity of the disorder and $\delta_{m,n}$ is the Kronecker $\delta$ function. The equilibrium zero temperature configuration was obtained using the transfer-matrix method~\cite{Kardar_prl_85_DP} with a droplet geometry, \textit{i.e.} with one end pinned at the origin while the other end is free.
Given the disordered potential ${V(u,z)}$, the weight ${Z(u,z)}$ of a polymer starting at ${(0,0)}$ and ending at ${(u,z)}$ is given recursively by
\begin{equation}
 Z(u,z) = e^{-\beta V(u,z)} \left[ Z(u-1,z-1) + Z(u+1,z-1) \right],
\end{equation}
with initial condition ${Z(u,0) = \delta_{u,0}}$, and $\beta$ the inverse temperature parameter. For each realization of the disordered potential the path of minimum energy corresponds to the largest weight, thus defining equilibrium interface~\cite{Kardar_prl_85_DP}.
In this canonical case, the value of the roughness exponent ${\zeta^{\mathrm{1D}}_{\mathrm{RB}} = 2/3}$ is already well known~\cite{huse_henley_fisher_1985_PhysRevLett55_2924}, allowing the accuracy of the roughness exponent estimation methods to be quantitatively assessed, and in particular the sensitivity to the finite size of the system to be tested. To this end, system sizes ${L=512}$, $1024$ and $2048$ sites were used, with $10^4$ different disorder realizations for each size.

\begin{figure}
\includegraphics[width=\columnwidth]{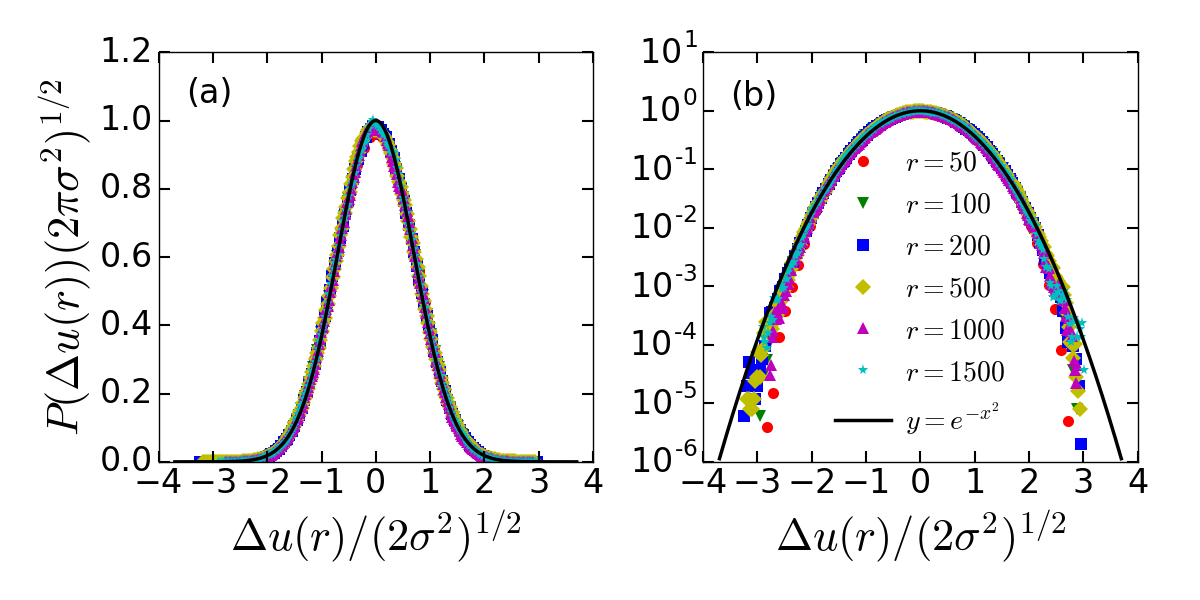}
\caption{
PDF of the relative displacements in Gaussian units for numerical interfaces in equilibrium, taken over $10^4$ numerical disorder configurations. The system size is ${L=2048}$ and different length scales $r$ are indicated in the key. ${\sigma = \sigma(r)}$, standing for the $r$-dependent standard deviation of the PDFs. The solid line corresponds to the Gaussian function. The same data is plotted in (a)~linear and (b)~lin-log scale in order to emphasize deviations from the Gaussian function.
}
\label{fig_num_equ_pdf_L}
\end{figure}

\begin{figure}
\includegraphics[width=\columnwidth]{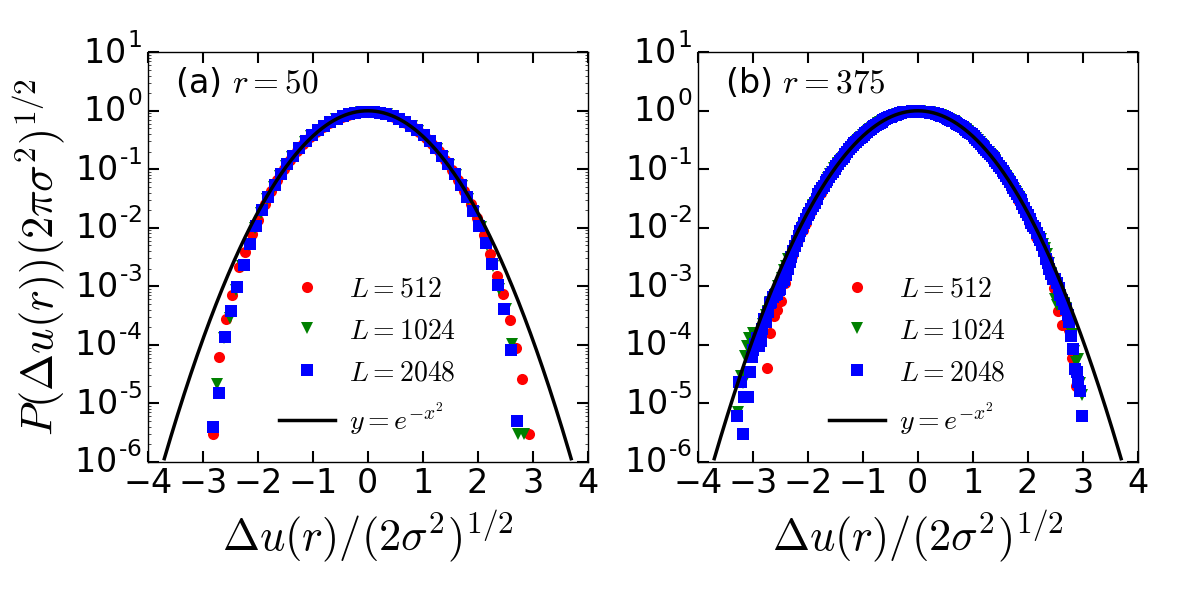}
\caption{
PDF of the relative displacements in Gaussian units for numerical interfaces in equilibrium, taken over $10^4$ numerical disorder configurations, for fixed values of the length scale $r$: (a)~${r = 50}$ and (b)~${r=375}$. Different system sizes are indicated in the keys.
}
\label{fig_num_equ_pdf_r}
\end{figure}

If they are mono-affine systems, these equilibrated interfaces should exhibit a PDF of relative displacements, ${P(\Delta u(r))}$, well described by a Gaussian function. Here ${\Delta u(r)}$ is the relative displacement between two points a distance $r$ apart (see Fig.~\ref{fig:def-roughness}).
As can be observed in \figref{fig_num_equ_pdf_L}, showing the displacements PDF for different length scales $r$ with system size ${L=2048}$, this is very well verified for intermediate length scales, \textit{i.e.} for ${100 \lesssim r \lesssim L/2}$.
For small $r$, slight finite-size effects, signaled by a weak finite-size dependence of the tail of the distribution, can be observed for all system sizes ($r=50$ in \figref{fig_num_equ_pdf_r}(a)).
As $r$ approaches the system size $L$, lack of statistics prevents sufficient averaging, as can be noticed in the tails of the distribution for ${r=375}$ in \figref{fig_num_equ_pdf_r}(b).
We note that these observations are similar to the behavior of the autocorrelation functions defined in Ref.~\cite{guyonnet_prl_12_multiscaling, santucci_pre_07_fracture_statistics}.

\begin{figure}[!htbp]
\includegraphics[width=\columnwidth]{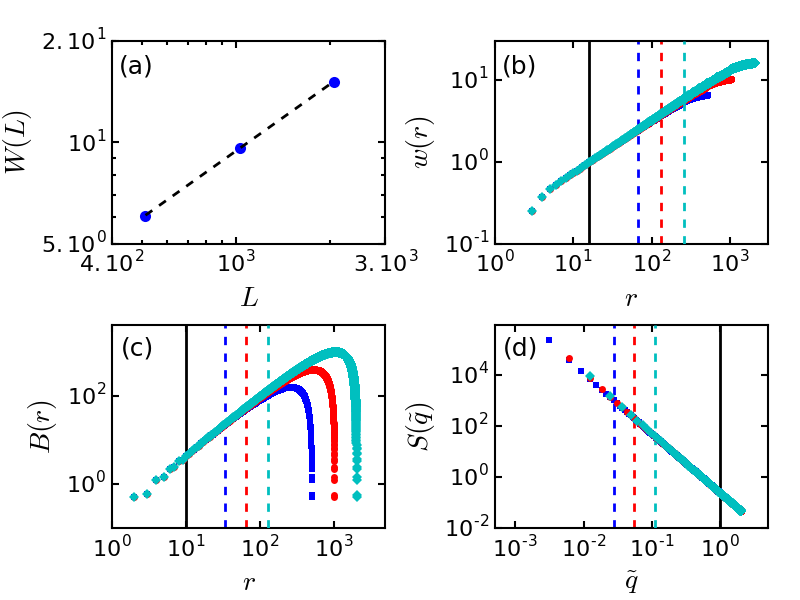}
\caption{Roughness analysis for numerical interfaces in equilibrium, averaged over $10^4$ disorder configurations:
(a)~global width ${W(L)}$, (b)~local width ${w(r)}$, (c)~displacement autocorrelation function ${B(r)}$ and (d)~structure factor ${S(q)}$.
In (b), (c), and (d) blue $\square$, red $\Circle$ and cyan $\Diamond$ correspond to ${L=512}$, $1024$ and $2048$, respectively.
The corresponding roughness exponents are listed in Table~\ref{table-roughness-equilibrium};
the solid vertical lines in (b)-(c)-(d) indicate the common lower bound of the fitting regions for the three sizes, while the upper bounds depend on the system size and are indicated by the vertical dashed lines.
\label{fig_num_equ_br_sq}
}
\end{figure}

\begin{table}[!th]
\begin{center}
\begin{tabular}{|c||c|c|c||}
\hline
$L$ & $512$	&	$1048$	&	$2048$ \\
\hline \hline
$\zeta_{W(L)}$	&	\multicolumn{3}{|c||}{${0.658 \pm 0.005}$}	\\

\hline
$\zeta_{w(r)}$ &	${0.633 \pm 0.001}$	&	${0.641 \pm 0.001}$	&	${0.646 \pm 0.001}$		\\
\hline
$\zeta_{B(r)}$	&	${0.655 \pm 0.006}$	&	${0.660 \pm 0.002}$	&	${0.660 \pm 0.001}$		\\
\hline
$\zeta_{S(q)}$  &	${0.659 \pm 0.001}$	&	${0.659 \pm 0.001}$	&	${0.658 \pm 0.001}$		\\
\hline
\end{tabular}
\end{center}
\caption{
Roughness exponent for 1D interfaces in equilibrium, obtained by fitting the graphs in Fig.~\ref{fig_num_equ_br_sq} over restricted length scale ranges.
Error bars come from the linear fits of the log-log plots in Fig.~\ref{fig_num_equ_br_sq}.
\label{table-roughness-equilibrium}
}
\end{table}

For the roughness exponent, the values obtained using the different roughness definitions are listed in Table~\ref{table-roughness-equilibrium}, with their corresponding error bars, and they are all close to the expected value of ${\zeta^{\mathrm{1D}}_{\mathrm{RB}} = 2/3}$ (error bars in Table~\ref{table-roughness-equilibrium} are given by the linear fit and are probably underestimated).
Although only three different sizes are considered, we obtain ${\zeta_{W(L)}=0.658}$ from the global width averaged over all disorder realizations, Fig.~\ref{fig_num_equ_br_sq}(a).
A value close to ${\zeta^{\mathrm{1D}}_{\mathrm{RB}}}$ is also recovered for ${L=2048}$ from the local width analysis, with ${\zeta_{w(r)}=0.646}$, and the displacements autocorrelation functions in real and reciprocal space, with ${\zeta_{B(r)}=0.660}$ and ${\zeta_{S(q)}=0.658}$ (\figref{fig_num_equ_br_sq}(b--d)).
When fitting the latter three quantities, similar size effects can be observed for very small and large length scales when considering ${1 \le r \le L-1}$, thus affecting the choice of adequate fitting ranges.
For the local width, this manifests as lower width values, resulting in loss of power-law behaviors on either ends of the length scale range: based on the best correlation coefficients, the best fits are obtained between ${r=16}$ and ${r \approx L/8}$, indicated by the vertical solid and 
dashed lines in \figref{fig_num_equ_br_sq}(b).
In the local width approach, systematic errors due to finite system size therefore appear to lead to slightly underestimated values of the roughness exponent.
We note here that for real systems presenting both experimental artifacts and fewer realizations for averaging, we expect this effect to be significantly greater.

As can be reasonably expected in the real space displacements autocorrelation functions ${B(r)}$, the power-law behavior observed at intermediate length scales breaks down from the loss of statistics around ${r=L/2}$, corresponding to the local maximum of ${B(r)}$ observed in \figref{fig_num_equ_br_sq}(c). However, the most adequate fitting ranges are found to extend between ${r=10}$ and ${r \approx L/16}$, again indicated by the vertical solid and dashed lines in \figref{fig_num_equ_br_sq}(c). Although the power-law behavior degrades faster as $L$ increases than it does for the local width, resulting in narrower fitting ranges, scaling exponents are consistently found to be ${\zeta_{B(r)} = 0.66}$ for all system sizes. We note that if the same narrower ranges are considered for the local width, no change in the values of the roughness exponents is observed, thus indicating that the slight better accuracy of the ${B(r)}$ method is not a direct consequence of different power-law fitting ranges.

Finally, the structure factor functions represented in \figref{fig_num_equ_br_sq}(d) show the least visible amount of finite-size effects of all three methods, with the slight increase at small $q$ mirroring the lowering trend at large $r$ in real space methods. Best fitting ranges were again chosen as a function of the regression correlation coefficient and found to lie between ${\tilde{q}=2\sin(8\pi/L)}$ and ${\tilde{q} = 1}$, yielding size-independent values of ${\zeta_{S(q)} = 0.66}$.

\begin{figure}
\centering
\includegraphics[width=\columnwidth]{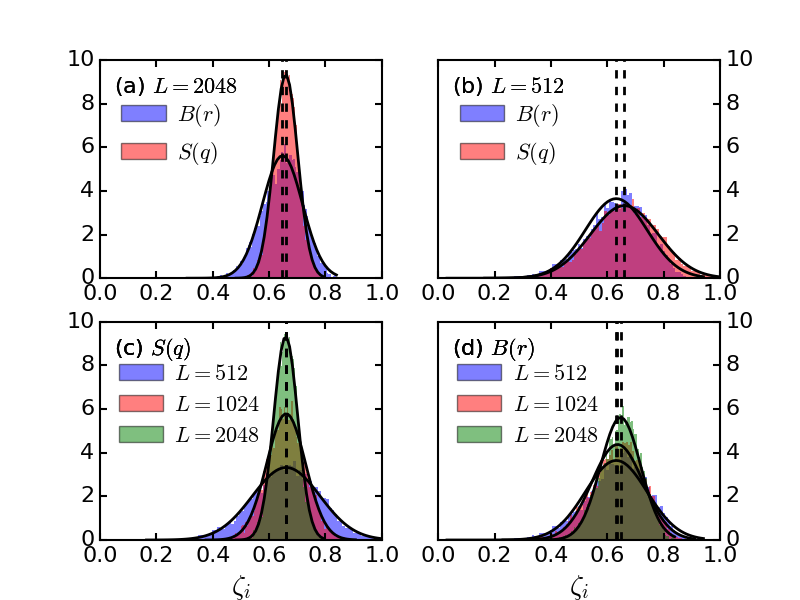}
\caption{Histograms of the roughness exponent obtained using ${B(r)}$ and ${S(q)}$ for individual numerical interfaces in equilibrium.
Comparing ${B(r)}$ and ${S(q)}$ at equivalent system sizes, (a)~${L=2048}$ and (b)~${L=512}$; wider histograms are obtained with ${B(r)}$.
Moreover, comparing finite-size effects for both (c)~${S(q)}$ and (d)~${B(r)}$, it can be observed that the histograms obtained with the structure factor converge more rapidly. Dashed lines indicate average values for each histogram.
}
\label{fig_num_equ_hist}
\end{figure}

\begin{figure}
\includegraphics[width=\columnwidth]{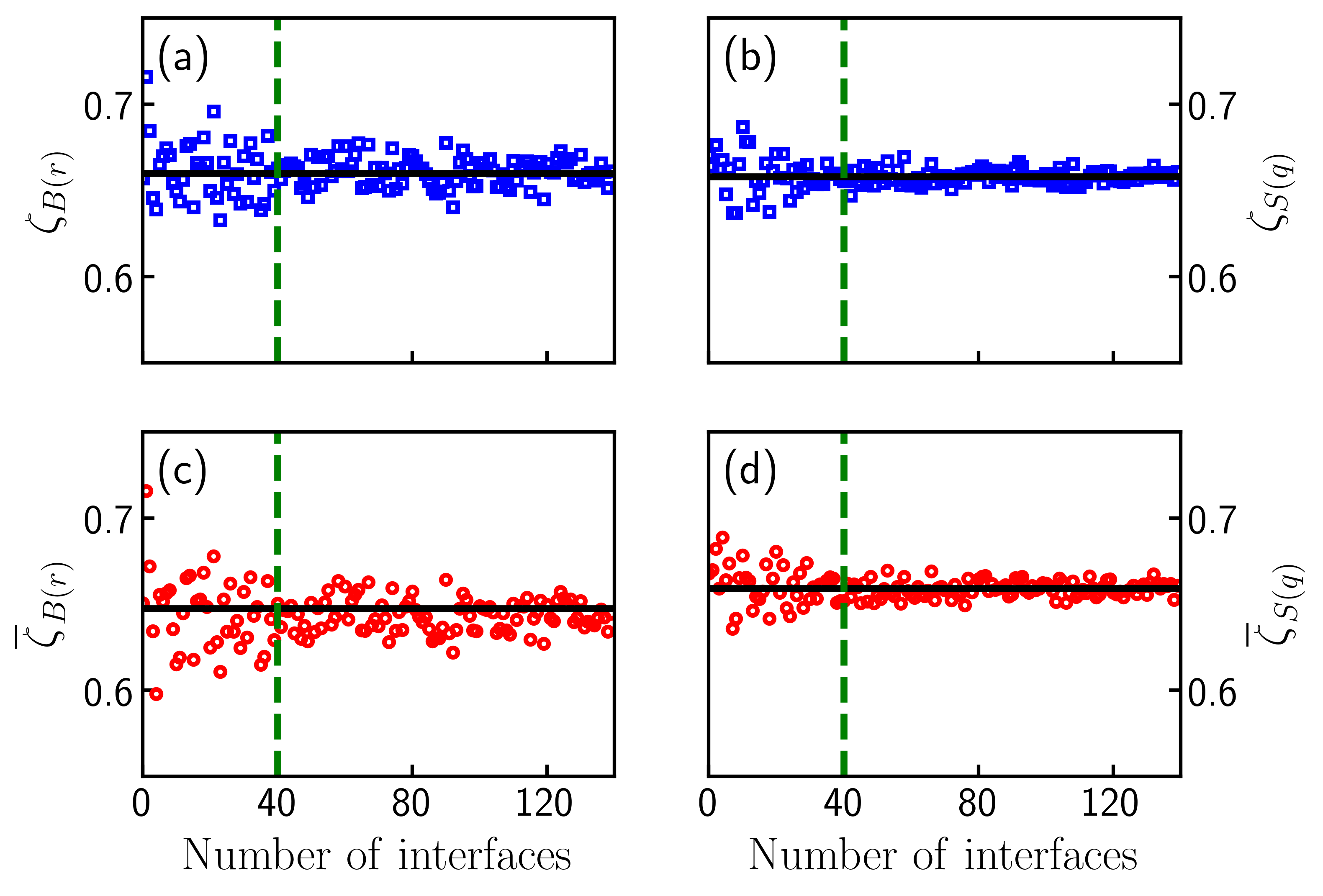}
\caption{
Comparison of the roughness exponents $\zeta$ (fit of the averaged quantities) and $\bar{\zeta}$ (average of the individual fits) for small numbers of independent disorder realizations, obtained either from ${B(r)}$ in (a,c), or from ${S(q)}$ in (b,d). Interfaces with ${L=2048}$ are used. Horizontal lines corresponds to the average values obtained for $10^4$ interfaces, as reported in Tables~\ref{table-roughness-equilibrium} and \ref{table-roughness-equilibrium-size}. Vertical lines indicate the number of interfaces needed to obtained a converged roughness exponent value, which is around 40 for all cases.
}
\label{fig_num_equ_fewsamples}
\end{figure}

At this point, it is therefore not clear which, if any, of the real-space or reciprocal-space autocorrelation function methods should be preferred. However, their different sensitivities to size effects, only marginally observable on quantities averaged over $10^4$ different disorder realizations, can be expected to become much more prominent in studies where disorder averaging is significantly reduced. This is immediately verified, as can be seen from the distributions of individual realization exponents extracted from ${B(r)}$ and ${S(q)}$, shown in \figref{fig_num_equ_hist}(a,b), as introduced in Sec.~\ref{sec_zeta_avg}.
In both cases, scaling exponents are obtained from power-law fits for each single interface, with the same fitting regions as the ones defined for the averaged quantities. For ${L=2048}$, the histogram of the individual exponent values constructed from ${B(r)}$ appears significantly wider than the one from ${S(q)}$, with full-width-at-half-maximum (FWHM) of 0.17 and 0.10 respectively.
In contrast, both methods yield histograms of comparable widths for ${L=512}$, suggesting the convergence of the distribution with increasing system size happens faster for the structure factor method, as shown in Table~\ref{table-roughness-equilibrium-size}.
It is worth noting here that the error bars of the mean values are small, reflecting the fact that a large number of interfaces ($10^4$) is available for statistical analysis. However, the large values of the FWHM indicate that the distributions are rather wide, and values of the roughness exponent for independent interfaces may differ considerably beyond the error bar of the mean. This fact should be taken into account when studying a small number of interface realizations, as is typically the case in experiments.
Another notable feature of the distributions presented in \figref{fig_num_equ_hist} is the slight negative skewness of all histograms, also decreasing with increasing system size, but significantly more pronounced for ${B(r)}$. This can be attributed to the inherent ${\zeta<1}$ cutoff of the method, as discussed in Sec.~\ref{sec_br_superrough}, effectively compressing the histogram to the right.
For ${S(q)}$, where this limitation is not present, the smaller skewness could originate from an intrinsic cutoff, namely the local solid-on-solid restriction of the model. Nevertheless, histograms computed from ${S(q)}$ are sufficiently symmetric to be well approximated by a Gaussian distribution (\figref{fig_num_equ_hist}(c)), which is reflected in the fact that the mean of all individual scaling exponents, denoted by $\overline{\zeta}$, agrees with the value of $0.66$ obtained previously from standard disorder averaging for all system sizes.
For ${B(r)}$ histograms, the small but non-negligible skewness is responsible for a slight underestimation of the roughness exponent (\figref{fig_num_equ_hist}(d)).

\begin{table}[!th]
\begin{center}
\begin{tabular}{|c||c|c|c||}
\hline
$L$ & $512$	&	$1024$	&	$2048$ \\
\hline \hline
$\bar{\zeta}_{B(r)}$	&	${0.632   \pm 0.001}$	&	${0.636 \pm 0.001}$	&	${0.6475 \pm 0.0007}$		\\
\hline
$\bar{\zeta}_{S(q)}$	&	${0.661 \pm 0.001}$	&	${0.6605 \pm 0.0007}$	&	${0.6590 \pm 0.0004}$		\\
\hline
\end{tabular}
\end{center}
\caption{
System size dependence of the roughness exponent for 1D interfaces in equilibrium. Notice that error bars of mean values are significantly smaller than FWHM values of the PDFs in \figref{fig_num_equ_hist}.
\label{table-roughness-equilibrium-size}
}
\end{table}

We address as a last issue regarding the distribution of roughness exponents the convergence of statistical averaging with the number of independent realizations. In order to do so, we simultaneously compute $\zeta_{B(r)}$, $\zeta_{S(q)}$, $\overline{\zeta}_{B(r)}$ and $\overline{\zeta}_{S(q)}$, with a small number of realizations ranging between 1 and 140, and no overlap between the averaging sets.
As can be observed in \figref{fig_num_equ_fewsamples} from the evolution of the roughness exponent values with the number of considered interfaces, the convergence to the average value (from Tables~\ref{table-roughness-equilibrium} and \ref{table-roughness-equilibrium-size}) is always obtained above of the order of 40 realizations, indicated by dashed vertical lines in the figure. Fluctuations appear large to the left of the vertical line and seems to be converged to its right.

In conclusion, the reciprocal-space autocorrelation method appears more accurate for the determination of the roughness exponent, with a particularly convenient equivalence between fitting the averaged correlation function and averaging individual exponents.
However, the real-space method based on the roughness function ${B(r)}$ could still be used to obtain a meaningful estimation of the roughness exponent for very short interfaces, as it might be the case for experimental realizations of ferroelectric or ferromagnetic domain walls.
Furthermore, the local width and the displacement auto-correlation function lead to a slight underestimation of the roughness exponents.
Finally, the averaging convergence results indicate that a minimum of a few tens of independent realizations should be considered in order to provide a meaningful roughness exponent analysis, regardless of the method which is chosen.

\subsection{Interfaces in critical depinning}
\label{sec_num_dep}

When an interface living in a disordered energy landscape is driven by an external force, its zero temperature critical depinning state corresponds to the configuration encountered exactly at the depinning force, separating zero velocity from finite velocity steady states~\cite{Ferrero2013}. This critical depinning state then results from the interplay between the elasticity of the interface, the disordered energy landscape and the external force.
We use a simple model to describe the dynamics of an elastic interface in a disordered energy landscape given by
\begin{eqnarray}
 \partial_t u(z,t) = & &u(z+1,t) - 2 u(z,t) + u(z-1,t) \nonumber \\
  &+& F_p\left( u(z,t),z \right) + F,
\end{eqnarray}
where ${u(z,t)}$ is the time-dependent position of the interface, $F$ is an homogeneous external drive, and ${F_p(u,z)= - \partial_u V(u,z)}$ is a pinning force. The disordered potential ${V(u,z)}$ has zero mean and correlations ${\overline{\left[ V(u,z)-V(u',z') \right]^2} = D \delta_{z,z'} R(u-u')}$, with ${R(u)}$ decaying in a finite range. For this model, there exists a finite force value $F_d$ separating pinning configurations for ${F < F_d}$ from moving configurations for ${F > F_d}$. $F_d$ is the pinning force and Middleton theorems~\cite{Middleton1992} assure that there exists a unique critical depinning configuration ${u_c(z)}$ corresponding to ${\partial_t u(z,t) = 0}$ at $F_d$. 
1D interfaces in a critical depinning state were obtained using the algorithm developed in Ref.~\cite{rosso_prl_03_depinning}, where the interface is forced to its last zero-velocity state under a finite driving force. The roughness exponent characterizing critical depinning interfaces is then ${\zeta_{\mathrm{dep,harm}}^{1D}=1.25}$ (see Ref.~\cite{Ferrero_pre_13_numerical_exponents}).
The simulation box longitudinal and transverse sizes $L$ and $M$ were chosen such that ${L/M \sim 3-10}$ in order to avoid spurious effects due to periodic boundary conditions~\cite{Bustingorry_prb_10_random_periodic,Bustingorry_pip_10_random_periodic,Kolton_jstat_13_fc_random_periodic}, and simulations were performed with ${L=64}$, $128$, $256$, $512$, and $1024$, with $10^3$ independent disorder configurations for each size. In this model, the internal 
coordinate of the interface position is a discrete variable ${z=1,2,\ldots,L}$ and the transverse coordinate 
by a continuous variable ${u(z)\in\mathbb{R}}$.

\begin{figure}[t]
\includegraphics[width=\columnwidth]{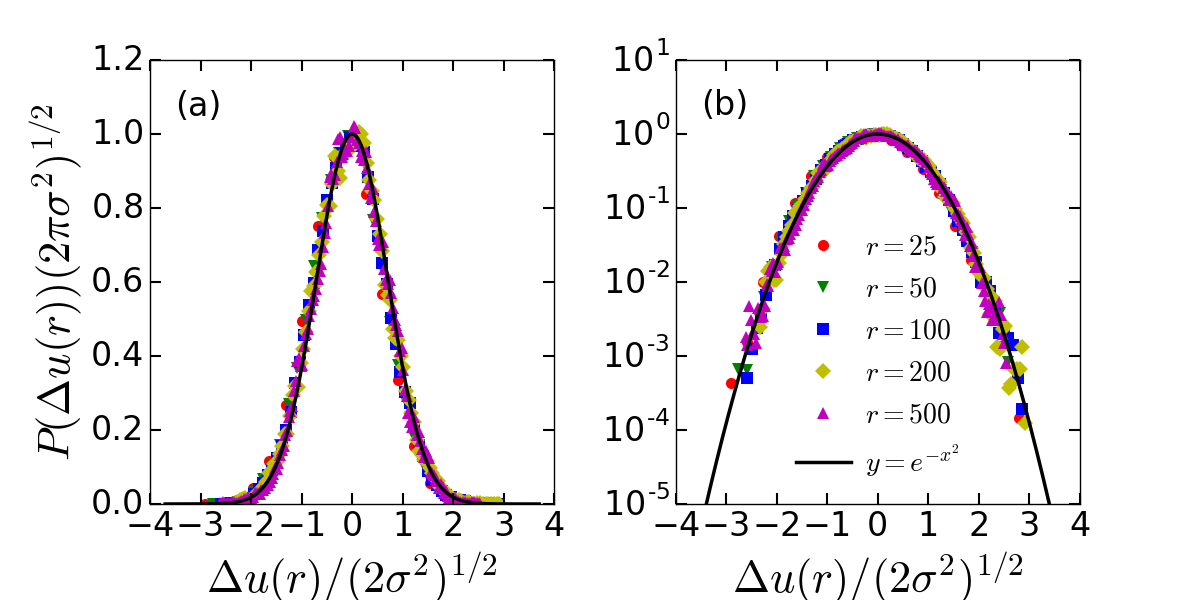}
\caption{PDF of the relative displacements in Gaussian units for numerical 
interfaces in a critical depinning state, taken over $10^3$ numerical disorder configurations. The system size is ${L=1024}$ and different length scales $r$ are indicated in the key. $\sigma$ stands for the standard deviation and the solid line is the Gaussian function. The same data is plotted in (a) linear and (b) lin-log scales.
}
\label{fig_num_dep_pdf}
\end{figure}

Here, we observe that the displacements PDFs at different length scales $r$, shown in \figref{fig_num_dep_pdf} for ${L=1024}$, are in excellent agreement with a Gaussian function at all length scales. Notably, there is no visible size effect for small $r$, in contrast with the equilibrium simulations in which small deviations are caused by the discretization of ${u(z)}$. Thus, the interfaces in critical depinning are geometrically mono-affine, and as we will see, they have a well-defined roughness exponent.

\begin{figure}[t]
\includegraphics[width=\columnwidth]{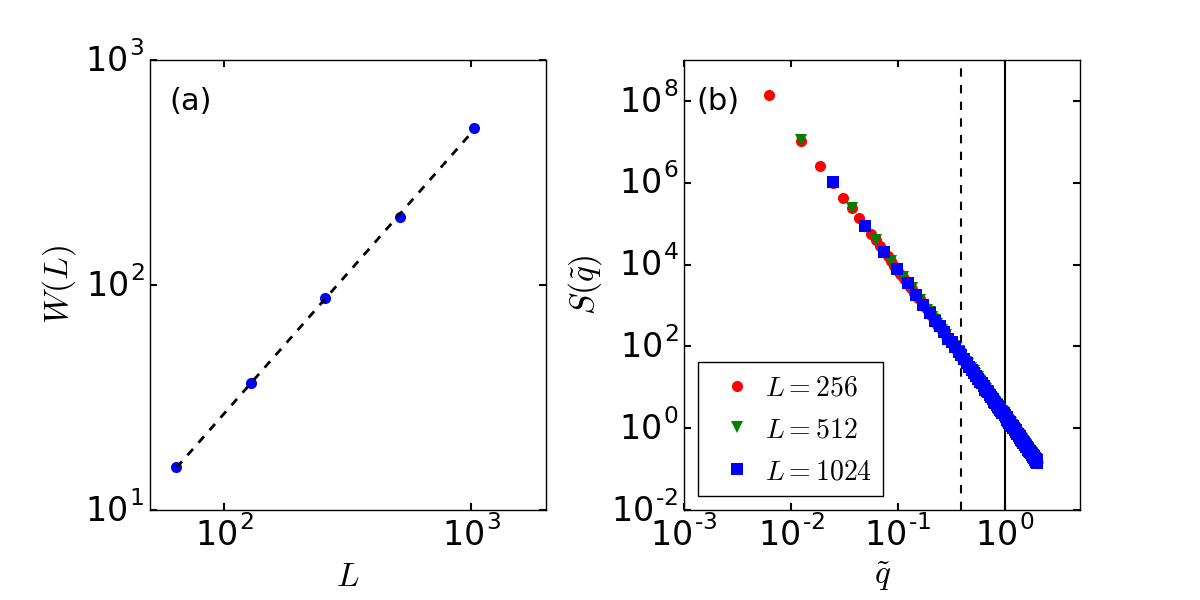}
\caption{Roughness analysis for numerical interfaces in a critical depinning state, averaged over 10$^3$ disorder configurations:
(a)~global width ${W(L)}$, (b)~structure factor ${S(q)}$.
The solid and dashed vertical lines in (b) indicate the lower and upper length scale bounds used to fit the power-law behavior.
The corresponding roughness exponents are listed in Table~\ref{table-roughness-depinning}.
}
\label{fig_num_dep_WL_sq}
\end{figure}

From the average global width as a function of $L$ (see \figref{fig_num_dep_WL_sq}(a)), we obtain ${\zeta_{W(L)} = 1.25}$, in good agreement with the expected value for the roughness exponent ${\zeta_{\mathrm{dep,harm}}^{1D}=1.25}$ (see Ref.\cite{Ferrero_pre_13_numerical_exponents}). This value is also recovered from the structure factor method, with the fitting range determined from the regression quality to be between ${\tilde q=1}$ and ${\tilde q \approx 0.39}$ for all system sizes (see \figref{fig_num_dep_WL_sq}(b)). The obtained values are listed in Table~\ref{table-roughness-depinning}.

\begin{table}[!th]
\begin{center}
\begin{tabular}{|c||c|c|c|}
\hline \hline
$L$ &	$256$	&	$512$	&	$1024$ \\
\hline \hline
$\zeta_{W(L)}$	&	\multicolumn{3}{c|}{${1.25 \pm 0.01}$}	\\
\hline
$\zeta_{S(q)}$  &	${1.261 \pm 0.04}$	&	${1.263 \pm 0.05}$	&	${1.27 \pm 0.01}$	\\
\hline \hline
$\zeta_{w(r)}$ &	${0.923 \pm 0.003}$	&	${0.932 \pm 0.001}$	&	${0.945 \pm 0.001}$	\\
\hline
$\zeta_{B(r)}$	&	${0.935 \pm 0.003}$	&	${0.942 \pm 0.002}$	&	${0.949 \pm 0.002}$	\\
\hline \hline
$\zeta_{B(r), SR}$	&	${1.2683 \pm 0.0007}$	&	${1.2713 \pm 0.0004}$	&	${1.2650 \pm 0.0003}$	\\
\hline
\end{tabular}
\end{center}
\caption{
Roughness exponent for 1D interfaces in a critical depinning state, obtained by fitting the graphs in Figs.~\ref{fig_num_dep_WL_sq}, \ref{fig_num_dep_wr_br} and \ref{fig_num_dep_br}: $\zeta_{W(L)}$ and $\zeta_{S(q)}$ are global roughness exponents while $\zeta_{w(r)}$ and $\zeta_{B(r)}$ gives the local value. Once the system-size dependent factor $C(L)$ is considered, the global value $\zeta_{B(r), SR}$ can be obtained from the ${B(r)}$ function. Error bars come from the linear fits of the corresponding log-log plots.
\label{table-roughness-depinning}
}
\end{table}

\begin{figure}[t]
\centering
\includegraphics[width=\columnwidth]{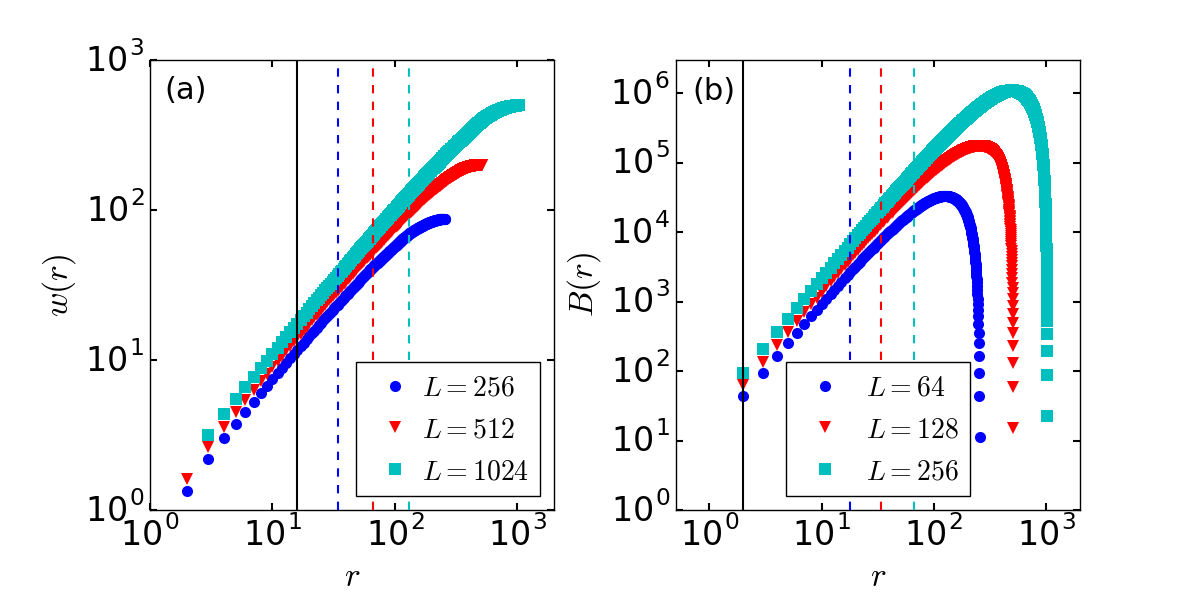}
\caption{(a)~Local widths and (b)~displacement autocorrelation functions computed for numerical interfaces in a critical depinning state, averaged over 10$^3$ disorder configurations, showing power-law trends characterized by scaling exponents close to $1$. The vertical offsets separating curves from different system size $L$ are a signature of the anomalous behavior due to ${\zeta>1}$.
The solid and dashed vertical lines indicate the lower and system size-dependent upper bounds used to fit the power-law behavior.
The obtained local roughness exponents for different system size are listed in Table.\ref{table-roughness-depinning}.
}
\label{fig_num_dep_wr_br}
\end{figure}

These values contrast noticeably with those obtained from the local width and real-space displacements autocorrelation function, with respectively ${\zeta_{w(r)}=0.945}$ and ${\zeta_{B(r)}=0.949}$ for the largest system size ${L=1024}$, see \figref{fig_num_dep_wr_br}.
As reported in Ref.~\cite{leschhorn_prl_93_superroughening}, this is a direct consequence of the fact that these methods are limited to ${\zeta \leq 1}$ by construction, or, following the approach in Ref.~\cite{ramasco_prl_00_generic_scaling}, that the local roughness exponent ${\zeta_{\mathrm{loc}}}$ never exceeds the value of 1. In fact, ${\zeta_{W(L)} = \zeta_{S(q)} > 1}$ obtained from the global width and structure factor should result in ${\zeta_{\mathrm{loc}}=1}$, which is in good agreement with our findings for $\zeta_{w(r)}$ and $\zeta_{B(r)}$. 
We note, however, that the same finite-size effects previously mentioned for the equilibrium simulations, \textit{i.e.} underestimating the exponent, affect the behavior of the local width, thus yielding a measured exponent lower than $1$.

\begin{figure}[!t]
\centering
\includegraphics[width=\columnwidth]{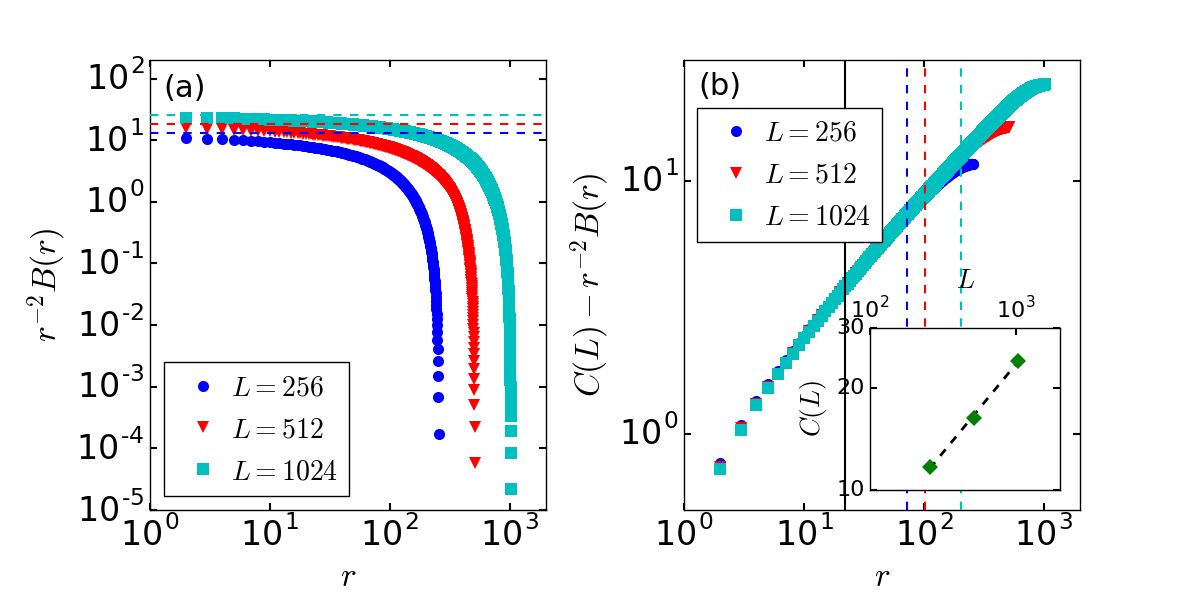}
\caption{Scaling of the displacement autocorrelation functions of super-rough numerical interfaces at critical depinning.
(a)~Estimation of the size-dependent constant ${C(L)}$, see~\eqref{equ_br_superrough}, from the $y$-intercept extrapolation of ${r^2B_{\mathrm{SR}}(r)}$ at small $r$, denoted by dashed lines.
(b)~Extraction of the roughness exponent, yielding ${\zeta_{B(r),\mathrm{SR}}(L=1024) = 1.2650}$, in excellent agreement with $\zeta_{W(L)}$ and $\zeta_{S(q)}$. Other values are reported in Table~\ref{table-roughness-depinning}.
Inset in (b) shows the scaling of the size-dependent constant ${C(L)}$, verified a posteriori, confirming ${\zeta_{B(r),\mathrm{SR}} = 1.26}$.
\label{fig_num_dep_br}
}
\end{figure}

As discussed in Sec.~\ref{sec_br_superrough} for super-rough interfaces with an exponent ${\zeta>1}$, there are three independent parameters in the mathematical expression of ${B_{\mathrm{SR}}(r)}$ given by \eqref{equ_br_superrough}: ${\left\lbrace C(L), A_1, A_2 \right \rbrace}$. 
The determination of the value of the roughness exponent cannot therefore be performed by a simple lest-square-fitting procedure.
However, this difficulty can in practice be overcome by estimating the ${C(L)}$ constant from the $y$-intercept extrapolated from ${r^2 B_{\mathrm{SR}}(r)}$ at small length scales $r$. The scaling of ${B_{\mathrm{SR}}(r)}$ is shown in \figref{fig_num_dep_br} for ${L=256}$, $512$, and $1024$. \figref{fig_num_dep_br}(a) shows the estimation of the $C(L)$ constants, where the ${r^2 B_{\mathrm{SR}}(r)}$ quantity is fitted over the first six points in order to extrapolate the $y$-intercept. Using these estimations, the roughness exponent can be extracted from the power-law behavior of ${C(L)-r^2 B_{\mathrm{SR}}(r)}$ plotted in \figref{fig_num_dep_br}(b).
Listed in Table~\ref{table-roughness-depinning}, the values of ${\zeta_{B(r),\mathrm{SR}}}$ are as expected in good agreement with the results from the global width and structure factor.
The consistency of this result and \eqref{equ_br_superrough} can be checked by verifying that the estimated $C(L)$ constants indeed obey ${C(L) \sim L^{2(\zeta-1)}}$, as shown in the inset of \figref{fig_num_dep_br}(b).

Thus, our results convincingly show that the real-space displacements autocorrelation function can be used to determine the global roughness exponent of super-rough interfaces, via the artifice of ${\zeta_{B(r),SR}}$.
As demonstrated for surface fractures, super-roughening has a significant impact on the morphology of the interface~\cite{lopez_pre_98_anomalous_scaling}.
In fact, two interfaces presenting an identical local roughness exponent (lower than $1$ by definition of ${B(r)}$ and ${w(r)}$) but different global roughness exponents (one of them being equal to the local roughness exponent, and the other greater than $1$) appear strikingly different.
However, a roughness scaling analysis assuming solely a Family-Vicsek behavior at all length scales would not allow this distinction to be made.
Since ${B(r)}$ is the easiest quantity to compute in experiments, \eqref{equ_br_superrough} provides a convenient way to assess the possibility of super-roughening (or more generally, with the appropriate ${B(r)}$ expression, of anomalous scaling) without having to compute the global width or structure factor.

\section{Summary}
\label{sec_concl}

In summary, we have compared different definitions of the roughness when applied to numerical 1D interfaces, showing that they correspond essentially to complementary approaches.
In the simplest case, they provide a cross-confirmation of the roughness exponent determination;
in the most complex cases, discrepancies between the $\zeta$ values obtained by either the real-space or the reciprocal-space approaches could signal anomalous behavior, such as super-roughening of the interface.
Moreover, such roughness analyses should be combined with a test of the mono-affinity of the interfaces, implemented via a `multiscaling' analysis, which can point out a possible breakdown of mono-affinity due to the presence of strong disorder pinning centers.

Our results reveal an important but often overlooked property of roughness characterization: The roughness exponent originates in wide size-dependent distributions. This should always be taken into account when evaluating the roughness exponent for a given problem. For instance, when reporting the roughness exponent, a number of the order of 50 independent realization of domain walls should be considered to guarantee statistical convergence to a meaningful average value. This result should prompt a reevaluation and development of detailed experimental protocols to assure statistical independence of domain wall configurations.
Such protocols would be particularly relevant for ferromagnetic and ferroelectric domain walls, since these experimental interfaces usually combine the issues of finite resolution, finite size, and limited number of experimental interfaces. However, our results are more broadly of interest for any experimental or numerical interfaces that could be described within the frame of disordered elastic systems.

\acknowledgements
The authors acknowledge enlightening discussions with J. Curiale, T. Giamarchi, M. Granada and A. B. Kolton.
JG and PP gratefully acknowledge financial support from the Fondation Ernst et Lucie Schmidheiny and SNSF grant 200021\_153174. EA acknowledges support from the SNF Ambizione Grant PZ00P2\_173962. SB acknowledges support from the Grant No. PIP11220120100250CO/CONICET from Argentina.


\begin{thebibliography}{62}%
\makeatletter
\providecommand \@ifxundefined [1]{%
 \@ifx{#1\undefined}
}%
\providecommand \@ifnum [1]{%
 \ifnum #1\expandafter \@firstoftwo
 \else \expandafter \@secondoftwo
 \fi
}%
\providecommand \@ifx [1]{%
 \ifx #1\expandafter \@firstoftwo
 \else \expandafter \@secondoftwo
 \fi
}%
\providecommand \natexlab [1]{#1}%
\providecommand \enquote  [1]{``#1''}%
\providecommand \bibnamefont  [1]{#1}%
\providecommand \bibfnamefont [1]{#1}%
\providecommand \citenamefont [1]{#1}%
\providecommand \href@noop [0]{\@secondoftwo}%
\providecommand \href [0]{\begingroup \@sanitize@url \@href}%
\providecommand \@href[1]{\@@startlink{#1}\@@href}%
\providecommand \@@href[1]{\endgroup#1\@@endlink}%
\providecommand \@sanitize@url [0]{\catcode `\\12\catcode `\$12\catcode
  `\&12\catcode `\#12\catcode `\^12\catcode `\_12\catcode `\%12\relax}%
\providecommand \@@startlink[1]{}%
\providecommand \@@endlink[0]{}%
\providecommand \url  [0]{\begingroup\@sanitize@url \@url }%
\providecommand \@url [1]{\endgroup\@href {#1}{\urlprefix }}%
\providecommand \urlprefix  [0]{URL }%
\providecommand \Eprint [0]{\href }%
\providecommand \doibase [0]{http://dx.doi.org/}%
\providecommand \selectlanguage [0]{\@gobble}%
\providecommand \bibinfo  [0]{\@secondoftwo}%
\providecommand \bibfield  [0]{\@secondoftwo}%
\providecommand \translation [1]{[#1]}%
\providecommand \BibitemOpen [0]{}%
\providecommand \bibitemStop [0]{}%
\providecommand \bibitemNoStop [0]{.\EOS\space}%
\providecommand \EOS [0]{\spacefactor3000\relax}%
\providecommand \BibitemShut  [1]{\csname bibitem#1\endcsname}%
\let\auto@bib@innerbib\@empty
\bibitem [{\citenamefont {Polla}\ and\ \citenamefont
  {Francis}(1998)}]{polla_97_mems}%
  \BibitemOpen
  \bibfield  {author} {\bibinfo {author} {\bibfnamefont {D.~L.}\ \bibnamefont
  {Polla}}\ and\ \bibinfo {author} {\bibfnamefont {L.~F.}\ \bibnamefont
  {Francis}},\ }\href {https://doi.org/10.1146/annurev.matsci.28.1.563}
  {\bibfield  {journal} {\bibinfo  {journal} {Annu. Rev. Mater. Sci.}\ }\textbf
  {\bibinfo {volume} {28}},\ \bibinfo {pages} {563} (\bibinfo {year}
  {1998})}\BibitemShut {NoStop}%
\bibitem [{\citenamefont {Kumar}\ \emph {et~al.}(2004)\citenamefont {Kumar},
  \citenamefont {Paruch}, \citenamefont {Triscone}, \citenamefont {Daniau},
  \citenamefont {Ballandras}, \citenamefont {Pellegrino}, \citenamefont
  {Marr{\'e}},\ and\ \citenamefont {Tybell}}]{kumar_apl_04_SAW}%
  \BibitemOpen
  \bibfield  {author} {\bibinfo {author} {\bibfnamefont {A.~K.~S.}\
  \bibnamefont {Kumar}}, \bibinfo {author} {\bibfnamefont {P.}~\bibnamefont
  {Paruch}}, \bibinfo {author} {\bibfnamefont {J.-M.}\ \bibnamefont
  {Triscone}}, \bibinfo {author} {\bibfnamefont {W.}~\bibnamefont {Daniau}},
  \bibinfo {author} {\bibfnamefont {S.}~\bibnamefont {Ballandras}}, \bibinfo
  {author} {\bibfnamefont {L.}~\bibnamefont {Pellegrino}}, \bibinfo {author}
  {\bibfnamefont {D.}~\bibnamefont {Marr{\'e}}}, \ and\ \bibinfo {author}
  {\bibfnamefont {T.}~\bibnamefont {Tybell}},\ }\href
  {http://scitation.aip.org/content/aip/journal/apl/85/10/10.1063/1.1787897}
  {\bibfield  {journal} {\bibinfo  {journal} {Appl. Phys. Lett.}\ }\textbf
  {\bibinfo {volume} {85}},\ \bibinfo {pages} {1757} (\bibinfo {year}
  {2004})}\BibitemShut {NoStop}%
\bibitem [{\citenamefont {Scott}\ and\ \citenamefont
  {de~Araujo}(1989)}]{scott_sci_89_memories}%
  \BibitemOpen
  \bibfield  {author} {\bibinfo {author} {\bibfnamefont {J.~F.}\ \bibnamefont
  {Scott}}\ and\ \bibinfo {author} {\bibfnamefont {C.~A.~P.}\ \bibnamefont
  {de~Araujo}},\ }\href {\doibase 10.1126/science.246.4936.1400} {\bibfield
  {journal} {\bibinfo  {journal} {Science}\ }\textbf {\bibinfo {volume}
  {246}},\ \bibinfo {pages} {1400} (\bibinfo {year} {1989})}\BibitemShut
  {NoStop}%
\bibitem [{\citenamefont {Waser}\ and\ \citenamefont
  {R{\"u}diger}(2004)}]{waser_natmat_04_memories}%
  \BibitemOpen
  \bibfield  {author} {\bibinfo {author} {\bibfnamefont {R.}~\bibnamefont
  {Waser}}\ and\ \bibinfo {author} {\bibfnamefont {A.}~\bibnamefont
  {R{\"u}diger}},\ }\href {\doibase 10.1038/nmat1067} {\bibfield  {journal}
  {\bibinfo  {journal} {Nat. Mater.}\ }\textbf {\bibinfo {volume} {3}},\
  \bibinfo {pages} {81} (\bibinfo {year} {2004})}\BibitemShut {NoStop}%
\bibitem [{\citenamefont {Parkin}\ \emph {et~al.}(2008)\citenamefont {Parkin},
  \citenamefont {Hayashi},\ and\ \citenamefont {Thomas}}]{Parkin2008}%
  \BibitemOpen
  \bibfield  {author} {\bibinfo {author} {\bibfnamefont {S.~S.~P.}\
  \bibnamefont {Parkin}}, \bibinfo {author} {\bibfnamefont {M.}~\bibnamefont
  {Hayashi}}, \ and\ \bibinfo {author} {\bibfnamefont {L.}~\bibnamefont
  {Thomas}},\ }\href {\doibase 10.1126/science.1145799} {\bibfield  {journal}
  {\bibinfo  {journal} {Science}\ }\textbf {\bibinfo {volume} {320}},\ \bibinfo
  {pages} {190} (\bibinfo {year} {2008})}\BibitemShut {NoStop}%
\bibitem [{\citenamefont {Hayashi}\ \emph {et~al.}(2008)\citenamefont
  {Hayashi}, \citenamefont {Thomas}, \citenamefont {Moriya}, \citenamefont
  {Rettner},\ and\ \citenamefont {Parkin}}]{Hayashi2008}%
  \BibitemOpen
  \bibfield  {author} {\bibinfo {author} {\bibfnamefont {M.}~\bibnamefont
  {Hayashi}}, \bibinfo {author} {\bibfnamefont {L.}~\bibnamefont {Thomas}},
  \bibinfo {author} {\bibfnamefont {R.}~\bibnamefont {Moriya}}, \bibinfo
  {author} {\bibfnamefont {C.}~\bibnamefont {Rettner}}, \ and\ \bibinfo
  {author} {\bibfnamefont {S.~S.~P.}\ \bibnamefont {Parkin}},\ }\href
  {https://science.sciencemag.org/content/320/5873/209} {\bibfield  {journal}
  {\bibinfo  {journal} {Science}\ }\textbf {\bibinfo {volume} {320}},\ \bibinfo
  {pages} {209} (\bibinfo {year} {2008})}\BibitemShut {NoStop}%
\bibitem [{\citenamefont {Allwood}\ \emph {et~al.}(2005)\citenamefont
  {Allwood}, \citenamefont {Xiong}, \citenamefont {Faulkner}, \citenamefont
  {Atkinson}, \citenamefont {Petit},\ and\ \citenamefont
  {Cowburn}}]{Allwood2005}%
  \BibitemOpen
  \bibfield  {author} {\bibinfo {author} {\bibfnamefont {D.~A.}\ \bibnamefont
  {Allwood}}, \bibinfo {author} {\bibfnamefont {G.}~\bibnamefont {Xiong}},
  \bibinfo {author} {\bibfnamefont {C.~C.}\ \bibnamefont {Faulkner}}, \bibinfo
  {author} {\bibfnamefont {D.}~\bibnamefont {Atkinson}}, \bibinfo {author}
  {\bibfnamefont {D.}~\bibnamefont {Petit}}, \ and\ \bibinfo {author}
  {\bibfnamefont {R.~P.}\ \bibnamefont {Cowburn}},\ }\href
  {https://science.sciencemag.org/content/309/5741/1688} {\bibfield  {journal}
  {\bibinfo  {journal} {Science}\ }\textbf {\bibinfo {volume} {309}},\ \bibinfo
  {pages} {1688} (\bibinfo {year} {2005})}\BibitemShut {NoStop}%
\bibitem [{\citenamefont {Lemerle}\ \emph {et~al.}(1998)\citenamefont
  {Lemerle}, \citenamefont {Ferr{\'e}}, \citenamefont {Chappert}, \citenamefont
  {Mathet}, \citenamefont {Giamarchi},\ and\ \citenamefont {{Le
  Doussal}}}]{lemerle_prl_98_FMDW_creep}%
  \BibitemOpen
  \bibfield  {author} {\bibinfo {author} {\bibfnamefont {S.}~\bibnamefont
  {Lemerle}}, \bibinfo {author} {\bibfnamefont {J.}~\bibnamefont {Ferr{\'e}}},
  \bibinfo {author} {\bibfnamefont {C.}~\bibnamefont {Chappert}}, \bibinfo
  {author} {\bibfnamefont {V.}~\bibnamefont {Mathet}}, \bibinfo {author}
  {\bibfnamefont {T.}~\bibnamefont {Giamarchi}}, \ and\ \bibinfo {author}
  {\bibfnamefont {P.}~\bibnamefont {{Le Doussal}}},\ }\href
  {http://link.aps.org/doi/10.1103/PhysRevLett.80.849} {\bibfield  {journal}
  {\bibinfo  {journal} {Phys. Rev. Lett.}\ }\textbf {\bibinfo {volume} {80}},\
  \bibinfo {pages} {849} (\bibinfo {year} {1998})}\BibitemShut {NoStop}%
\bibitem [{\citenamefont {Paruch}\ \emph {et~al.}(2006)\citenamefont {Paruch},
  \citenamefont {Giamarchi}, \citenamefont {Tybell},\ and\ \citenamefont
  {Triscone}}]{paruch_jap_06_dynamics_FE}%
  \BibitemOpen
  \bibfield  {author} {\bibinfo {author} {\bibfnamefont {P.}~\bibnamefont
  {Paruch}}, \bibinfo {author} {\bibfnamefont {T.}~\bibnamefont {Giamarchi}},
  \bibinfo {author} {\bibfnamefont {T.}~\bibnamefont {Tybell}}, \ and\ \bibinfo
  {author} {\bibfnamefont {J.-M.}\ \bibnamefont {Triscone}},\ }\href
  {https://aip.scitation.org/doi/10.1063/1.2337356} {\bibfield  {journal}
  {\bibinfo  {journal} {J. Appl. Phys.}\ }\textbf {\bibinfo {volume} {100}},\
  \bibinfo {pages} {051608} (\bibinfo {year} {2006})}\BibitemShut {NoStop}%
\bibitem [{\citenamefont {Paruch}\ \emph {et~al.}(2007)\citenamefont {Paruch},
  \citenamefont {Giamarchi},\ and\ \citenamefont
  {Triscone}}]{paruch_dw_review_07}%
  \BibitemOpen
  \bibfield  {author} {\bibinfo {author} {\bibfnamefont {P.}~\bibnamefont
  {Paruch}}, \bibinfo {author} {\bibfnamefont {T.}~\bibnamefont {Giamarchi}}, \
  and\ \bibinfo {author} {\bibfnamefont {J.-M.}\ \bibnamefont {Triscone}},\
  }\href@noop {} {\emph {\bibinfo {title} {Physics of Ferroelectrics, a Modern
  Perspective}}},\ edited by\ \bibinfo {editor} {\bibfnamefont
  {K.}~\bibnamefont {Rabe}}, \bibinfo {editor} {\bibfnamefont {C.~H.}\
  \bibnamefont {Ahn}}, \ and\ \bibinfo {editor} {\bibfnamefont {J.-M.}\
  \bibnamefont {Triscone}}\ (\bibinfo  {publisher} {Springer},\ \bibinfo
  {address} {Berlin/Heidelberg},\ \bibinfo {year} {2007})\ p.\ \bibinfo {pages}
  {339}\BibitemShut {NoStop}%
\bibitem [{\citenamefont {Metaxas}\ \emph {et~al.}(2007)\citenamefont
  {Metaxas}, \citenamefont {Jamet}, \citenamefont {Mougin}, \citenamefont
  {Cormier}, \citenamefont {Ferr{\'e}}, \citenamefont {Baltz}, \citenamefont
  {Rodmacq}, \citenamefont {Dieny},\ and\ \citenamefont
  {Stamps}}]{metaxas_depinning_thermal_rounding}%
  \BibitemOpen
  \bibfield  {author} {\bibinfo {author} {\bibfnamefont {P.~J.}\ \bibnamefont
  {Metaxas}}, \bibinfo {author} {\bibfnamefont {J.~P.}\ \bibnamefont {Jamet}},
  \bibinfo {author} {\bibfnamefont {A.}~\bibnamefont {Mougin}}, \bibinfo
  {author} {\bibfnamefont {M.}~\bibnamefont {Cormier}}, \bibinfo {author}
  {\bibfnamefont {J.}~\bibnamefont {Ferr{\'e}}}, \bibinfo {author}
  {\bibfnamefont {V.}~\bibnamefont {Baltz}}, \bibinfo {author} {\bibfnamefont
  {B.}~\bibnamefont {Rodmacq}}, \bibinfo {author} {\bibfnamefont
  {B.}~\bibnamefont {Dieny}}, \ and\ \bibinfo {author} {\bibfnamefont {R.~L.}\
  \bibnamefont {Stamps}},\ }\href
  {http://link.aps.org/abstract/PRL/v99/e217208} {\bibfield  {journal}
  {\bibinfo  {journal} {Phys. Rev. Lett.}\ }\textbf {\bibinfo {volume} {99}},\
  \bibinfo {pages} {217208} (\bibinfo {year} {2007})}\BibitemShut {NoStop}%
\bibitem [{\citenamefont {Metaxas}\ \emph {et~al.}(2010)\citenamefont
  {Metaxas}, \citenamefont {Stamps}, \citenamefont {Jamet}, \citenamefont
  {Ferr{\'e}}, \citenamefont {Baltz}, \citenamefont {Rodmacq},\ and\
  \citenamefont {Politi}}]{Metaxas2010}%
  \BibitemOpen
  \bibfield  {author} {\bibinfo {author} {\bibfnamefont {P.~J.}\ \bibnamefont
  {Metaxas}}, \bibinfo {author} {\bibfnamefont {R.~L.}\ \bibnamefont {Stamps}},
  \bibinfo {author} {\bibfnamefont {J.-P.}\ \bibnamefont {Jamet}}, \bibinfo
  {author} {\bibfnamefont {J.}~\bibnamefont {Ferr{\'e}}}, \bibinfo {author}
  {\bibfnamefont {V.}~\bibnamefont {Baltz}}, \bibinfo {author} {\bibfnamefont
  {B.}~\bibnamefont {Rodmacq}}, \ and\ \bibinfo {author} {\bibfnamefont
  {P.}~\bibnamefont {Politi}},\ }\href
  {https://link.aps.org/doi/10.1103/PhysRevLett.104.237206} {\bibfield
  {journal} {\bibinfo  {journal} {Phys. Rev. Lett.}\ }\textbf {\bibinfo
  {volume} {104}},\ \bibinfo {pages} {237206} (\bibinfo {year}
  {2010})}\BibitemShut {NoStop}%
\bibitem [{\citenamefont {Giamarchi}\ \emph {et~al.}(2006)\citenamefont
  {Giamarchi}, \citenamefont {Kolton},\ and\ \citenamefont
  {Rosso}}]{giamarchi_domainwall_review}%
  \BibitemOpen
  \bibfield  {author} {\bibinfo {author} {\bibfnamefont {T.}~\bibnamefont
  {Giamarchi}}, \bibinfo {author} {\bibfnamefont {A.~B.}\ \bibnamefont
  {Kolton}}, \ and\ \bibinfo {author} {\bibfnamefont {A.}~\bibnamefont
  {Rosso}},\ }in\ \href@noop {} {\emph {\bibinfo {booktitle} {Jamming, Yielding
  and Irreversible deformation in condensed matter}}},\ \bibinfo {editor}
  {edited by\ \bibinfo {editor} {\bibfnamefont {M.~C.}\ \bibnamefont {Miguel}}\
  and\ \bibinfo {editor} {\bibfnamefont {J.~M.}\ \bibnamefont {Rubi}}}\
  (\bibinfo  {publisher} {Springer},\ \bibinfo {address} {Berlin/Heidelberg},\
  \bibinfo {year} {2006})\ p.~\bibinfo {pages} {91}\BibitemShut {NoStop}%
\bibitem [{\citenamefont {Barabasi}\ and\ \citenamefont
  {Stanley}(1995)}]{barabasi_surface_growth_95}%
  \BibitemOpen
  \bibfield  {author} {\bibinfo {author} {\bibfnamefont {A.-L.}\ \bibnamefont
  {Barabasi}}\ and\ \bibinfo {author} {\bibfnamefont {H.~E.}\ \bibnamefont
  {Stanley}},\ }\href@noop {} {\emph {\bibinfo {title} {Fractal Concepts in
  Surface Growth}}}\ (\bibinfo  {publisher} {Cambridge University Press},\
  \bibinfo {address} {New York, USA},\ \bibinfo {year} {1995})\BibitemShut
  {NoStop}%
\bibitem [{\citenamefont {Mandelbrot}\ \emph {et~al.}(1984)\citenamefont
  {Mandelbrot}, \citenamefont {Passoja},\ and\ \citenamefont
  {Paullay}}]{mandelbrot_nature_84_cracks_metal}%
  \BibitemOpen
  \bibfield  {author} {\bibinfo {author} {\bibfnamefont {B.~B.}\ \bibnamefont
  {Mandelbrot}}, \bibinfo {author} {\bibfnamefont {D.~E.}\ \bibnamefont
  {Passoja}}, \ and\ \bibinfo {author} {\bibfnamefont {A.~J.}\ \bibnamefont
  {Paullay}},\ }\href {https://doi.org/10.1038/308721a0} {\bibfield  {journal}
  {\bibinfo  {journal} {Nature}\ }\textbf {\bibinfo {volume} {308}},\ \bibinfo
  {pages} {721} (\bibinfo {year} {1984})}\BibitemShut {NoStop}%
\bibitem [{\citenamefont {Myllys}\ \emph {et~al.}(2000)\citenamefont {Myllys},
  \citenamefont {Maunuksela}, \citenamefont {Alava}, \citenamefont
  {Ala-Nissila},\ and\ \citenamefont {Timonen}}]{myllys_prl_00_burning_fronts}%
  \BibitemOpen
  \bibfield  {author} {\bibinfo {author} {\bibfnamefont {M.}~\bibnamefont
  {Myllys}}, \bibinfo {author} {\bibfnamefont {J.}~\bibnamefont {Maunuksela}},
  \bibinfo {author} {\bibfnamefont {M.~J.}\ \bibnamefont {Alava}}, \bibinfo
  {author} {\bibfnamefont {T.}~\bibnamefont {Ala-Nissila}}, \ and\ \bibinfo
  {author} {\bibfnamefont {J.}~\bibnamefont {Timonen}},\ }\href
  {https://link.aps.org/doi/10.1103/PhysRevLett.84.1946} {\bibfield  {journal}
  {\bibinfo  {journal} {Phys. Rev. Lett.}\ }\textbf {\bibinfo {volume} {84}},\
  \bibinfo {pages} {1946} (\bibinfo {year} {2000})}\BibitemShut {NoStop}%
\bibitem [{\citenamefont {Rubio}\ \emph {et~al.}(1989)\citenamefont {Rubio},
  \citenamefont {Edwards}, \citenamefont {Dougherty},\ and\ \citenamefont
  {Gollub}}]{rubio_prl_89_wetting_fronts}%
  \BibitemOpen
  \bibfield  {author} {\bibinfo {author} {\bibfnamefont {M.}~\bibnamefont
  {Rubio}}, \bibinfo {author} {\bibfnamefont {C.}~\bibnamefont {Edwards}},
  \bibinfo {author} {\bibfnamefont {A.}~\bibnamefont {Dougherty}}, \ and\
  \bibinfo {author} {\bibfnamefont {J.}~\bibnamefont {Gollub}},\ }\href
  {https://link.aps.org/doi/10.1103/PhysRevLett.63.1685} {\bibfield  {journal}
  {\bibinfo  {journal} {Phys. Rev. Lett.}\ }\textbf {\bibinfo {volume} {63}},\
  \bibinfo {pages} {1685} (\bibinfo {year} {1989})}\BibitemShut {NoStop}%
\bibitem [{\citenamefont {Bonachela}\ \emph {et~al.}(2011)\citenamefont
  {Bonachela}, \citenamefont {Nadell}, \citenamefont {Xavier},\ and\
  \citenamefont {Levin}}]{bonachela_jstatphys_11_bacterias_DES}%
  \BibitemOpen
  \bibfield  {author} {\bibinfo {author} {\bibfnamefont {J.~A.}\ \bibnamefont
  {Bonachela}}, \bibinfo {author} {\bibfnamefont {C.~D.}\ \bibnamefont
  {Nadell}}, \bibinfo {author} {\bibfnamefont {J.~B.}\ \bibnamefont {Xavier}},
  \ and\ \bibinfo {author} {\bibfnamefont {S.~A.}\ \bibnamefont {Levin}},\
  }\href {https://doi.org/10.1007/s10955-011-0179-x} {\bibfield  {journal}
  {\bibinfo  {journal} {J. Stat. Phys.}\ }\textbf {\bibinfo {volume} {144}},\
  \bibinfo {pages} {303} (\bibinfo {year} {2011})}\BibitemShut {NoStop}%
\bibitem [{\citenamefont {Chepizhko}\ \emph {et~al.}(2016)\citenamefont
  {Chepizhko}, \citenamefont {Giampietro}, \citenamefont {Mastrapasqua},
  \citenamefont {Nourazar}, \citenamefont {Ascagni}, \citenamefont {Sugni},
  \citenamefont {Fascio}, \citenamefont {Leggio}, \citenamefont {Malinverno},
  \citenamefont {Scita}, \citenamefont {Santucci}, \citenamefont {Alava},
  \citenamefont {Zapperi},\ and\ \citenamefont
  {La~Porta}}]{Chepizhko_pnas_16_cell_front}%
  \BibitemOpen
  \bibfield  {author} {\bibinfo {author} {\bibfnamefont {O.}~\bibnamefont
  {Chepizhko}}, \bibinfo {author} {\bibfnamefont {C.}~\bibnamefont
  {Giampietro}}, \bibinfo {author} {\bibfnamefont {E.}~\bibnamefont
  {Mastrapasqua}}, \bibinfo {author} {\bibfnamefont {M.}~\bibnamefont
  {Nourazar}}, \bibinfo {author} {\bibfnamefont {M.}~\bibnamefont {Ascagni}},
  \bibinfo {author} {\bibfnamefont {M.}~\bibnamefont {Sugni}}, \bibinfo
  {author} {\bibfnamefont {U.}~\bibnamefont {Fascio}}, \bibinfo {author}
  {\bibfnamefont {L.}~\bibnamefont {Leggio}}, \bibinfo {author} {\bibfnamefont
  {C.}~\bibnamefont {Malinverno}}, \bibinfo {author} {\bibfnamefont
  {G.}~\bibnamefont {Scita}}, \bibinfo {author} {\bibfnamefont
  {S.}~\bibnamefont {Santucci}}, \bibinfo {author} {\bibfnamefont {M.~J.}\
  \bibnamefont {Alava}}, \bibinfo {author} {\bibfnamefont {S.}~\bibnamefont
  {Zapperi}}, \ and\ \bibinfo {author} {\bibfnamefont {C.~A.~M.}\ \bibnamefont
  {La~Porta}},\ }\href {\doibase 10.1073/pnas.1600503113} {\bibfield  {journal}
  {\bibinfo  {journal} {PNAS}\ }\textbf {\bibinfo {volume} {113}},\ \bibinfo
  {pages} {11408} (\bibinfo {year} {2016})}\BibitemShut {NoStop}%
\bibitem [{\citenamefont {Speck}\ and\ \citenamefont
  {Vink}(2012)}]{speck_pre_12_cell_membrane}%
  \BibitemOpen
  \bibfield  {author} {\bibinfo {author} {\bibfnamefont {T.}~\bibnamefont
  {Speck}}\ and\ \bibinfo {author} {\bibfnamefont {R.~L.~C.}\ \bibnamefont
  {Vink}},\ }\href {https://link.aps.org/doi/10.1103/PhysRevE.86.031923}
  {\bibfield  {journal} {\bibinfo  {journal} {Phys. Rev. E}\ }\textbf {\bibinfo
  {volume} {86}},\ \bibinfo {pages} {031923} (\bibinfo {year}
  {2012})}\BibitemShut {NoStop}%
\bibitem [{\citenamefont {Paruch}\ and\ \citenamefont
  {Guyonnet}(2013)}]{paruch_13_FE_DW_review}%
  \BibitemOpen
  \bibfield  {author} {\bibinfo {author} {\bibfnamefont {P.}~\bibnamefont
  {Paruch}}\ and\ \bibinfo {author} {\bibfnamefont {J.}~\bibnamefont
  {Guyonnet}},\ }\href
  {http://www.sciencedirect.com/science/article/pii/S1631070513001321}
  {\bibfield  {journal} {\bibinfo  {journal} {C. R. Phys.}\ }\textbf {\bibinfo
  {volume} {14}},\ \bibinfo {pages} {667} (\bibinfo {year} {2013})}\BibitemShut
  {NoStop}%
\bibitem [{\citenamefont {Ferr\'e}\ \emph {et~al.}(2013)\citenamefont
  {Ferr\'e}, \citenamefont {Metaxas}, \citenamefont {Mougin}, \citenamefont
  {Jamet}, \citenamefont {Gorchon},\ and\ \citenamefont
  {Jeudy}}]{Ferre_cras_13_DW}%
  \BibitemOpen
  \bibfield  {author} {\bibinfo {author} {\bibfnamefont {J.}~\bibnamefont
  {Ferr\'e}}, \bibinfo {author} {\bibfnamefont {P.~J.}\ \bibnamefont
  {Metaxas}}, \bibinfo {author} {\bibfnamefont {A.}~\bibnamefont {Mougin}},
  \bibinfo {author} {\bibfnamefont {J.-P.}\ \bibnamefont {Jamet}}, \bibinfo
  {author} {\bibfnamefont {J.}~\bibnamefont {Gorchon}}, \ and\ \bibinfo
  {author} {\bibfnamefont {V.}~\bibnamefont {Jeudy}},\ }\href
  {http://www.sciencedirect.com/science/article/pii/S1631070513001291}
  {\bibfield  {journal} {\bibinfo  {journal} {C. R. Physique}\ }\textbf
  {\bibinfo {volume} {14}},\ \bibinfo {pages} {651} (\bibinfo {year}
  {2013})}\BibitemShut {NoStop}%
\bibitem [{\citenamefont {Agoritsas}\ \emph {et~al.}(2012)\citenamefont
  {Agoritsas}, \citenamefont {Lecomte},\ and\ \citenamefont
  {Giamarchi}}]{agoritsas_physb_12_DES}%
  \BibitemOpen
  \bibfield  {author} {\bibinfo {author} {\bibfnamefont {E.}~\bibnamefont
  {Agoritsas}}, \bibinfo {author} {\bibfnamefont {V.}~\bibnamefont {Lecomte}},
  \ and\ \bibinfo {author} {\bibfnamefont {T.}~\bibnamefont {Giamarchi}},\
  }\href {http://www.sciencedirect.com/science/article/pii/S0921452612000221}
  {\bibfield  {journal} {\bibinfo  {journal} {Physica B}\ }\textbf {\bibinfo
  {volume} {407}},\ \bibinfo {pages} {1725} (\bibinfo {year}
  {2012})}\BibitemShut {NoStop}%
\bibitem [{\citenamefont {Paruch}\ \emph {et~al.}(2005)\citenamefont {Paruch},
  \citenamefont {Giamarchi},\ and\ \citenamefont
  {Triscone}}]{paruch_prl_05_dw_roughness_FE}%
  \BibitemOpen
  \bibfield  {author} {\bibinfo {author} {\bibfnamefont {P.}~\bibnamefont
  {Paruch}}, \bibinfo {author} {\bibfnamefont {T.}~\bibnamefont {Giamarchi}}, \
  and\ \bibinfo {author} {\bibfnamefont {J.-M.}\ \bibnamefont {Triscone}},\
  }\href {http://link.aps.org/doi/10.1103/PhysRevLett.94.197601} {\bibfield
  {journal} {\bibinfo  {journal} {Phys. Rev. Lett.}\ }\textbf {\bibinfo
  {volume} {94}},\ \bibinfo {pages} {197601} (\bibinfo {year}
  {2005})}\BibitemShut {NoStop}%
\bibitem [{\citenamefont {Bauer}\ \emph {et~al.}(2005)\citenamefont {Bauer},
  \citenamefont {Mougin}, \citenamefont {Jamet}, \citenamefont {Repain},
  \citenamefont {Ferr\'e}, \citenamefont {Stamps}, \citenamefont {Bernas},\
  and\ \citenamefont {Chappert}}]{Bauer2005}%
  \BibitemOpen
  \bibfield  {author} {\bibinfo {author} {\bibfnamefont {M.}~\bibnamefont
  {Bauer}}, \bibinfo {author} {\bibfnamefont {A.}~\bibnamefont {Mougin}},
  \bibinfo {author} {\bibfnamefont {J.~P.}\ \bibnamefont {Jamet}}, \bibinfo
  {author} {\bibfnamefont {V.}~\bibnamefont {Repain}}, \bibinfo {author}
  {\bibfnamefont {J.}~\bibnamefont {Ferr\'e}}, \bibinfo {author} {\bibfnamefont
  {R.~L.}\ \bibnamefont {Stamps}}, \bibinfo {author} {\bibfnamefont
  {H.}~\bibnamefont {Bernas}}, \ and\ \bibinfo {author} {\bibfnamefont
  {C.}~\bibnamefont {Chappert}},\ }\href
  {https://link.aps.org/doi/10.1103/PhysRevLett.94.207211} {\bibfield
  {journal} {\bibinfo  {journal} {Phys. Rev. Lett.}\ }\textbf {\bibinfo
  {volume} {94}},\ \bibinfo {pages} {207211} (\bibinfo {year}
  {2005})}\BibitemShut {NoStop}%
\bibitem [{\citenamefont {Catalan}\ \emph {et~al.}(2008)\citenamefont
  {Catalan}, \citenamefont {B\'ea}, \citenamefont {Fusil}, \citenamefont
  {Bibes}, \citenamefont {Paruch}, \citenamefont {Barth\'el\'emy},\ and\
  \citenamefont {Scott}}]{catalan_prl_08_BFO_DW}%
  \BibitemOpen
  \bibfield  {author} {\bibinfo {author} {\bibfnamefont {G.}~\bibnamefont
  {Catalan}}, \bibinfo {author} {\bibfnamefont {H.}~\bibnamefont {B\'ea}},
  \bibinfo {author} {\bibfnamefont {S.}~\bibnamefont {Fusil}}, \bibinfo
  {author} {\bibfnamefont {M.}~\bibnamefont {Bibes}}, \bibinfo {author}
  {\bibfnamefont {P.}~\bibnamefont {Paruch}}, \bibinfo {author} {\bibfnamefont
  {A.}~\bibnamefont {Barth\'el\'emy}}, \ and\ \bibinfo {author} {\bibfnamefont
  {J.~F.}\ \bibnamefont {Scott}},\ }\href
  {http://link.aps.org/doi/10.1103/PhysRevLett.100.027602} {\bibfield
  {journal} {\bibinfo  {journal} {Phys. Rev. Lett.}\ }\textbf {\bibinfo
  {volume} {100}},\ \bibinfo {pages} {027602} (\bibinfo {year}
  {2008})}\BibitemShut {NoStop}%
\bibitem [{\citenamefont {Pertsev}\ \emph {et~al.}(2011)\citenamefont
  {Pertsev}, \citenamefont {Kiselev}, \citenamefont {Bdikin}, \citenamefont
  {Kosec},\ and\ \citenamefont {Kholkin}}]{pertsev_jap_11_ceramics}%
  \BibitemOpen
  \bibfield  {author} {\bibinfo {author} {\bibfnamefont {N.~A.}\ \bibnamefont
  {Pertsev}}, \bibinfo {author} {\bibfnamefont {D.~A.}\ \bibnamefont
  {Kiselev}}, \bibinfo {author} {\bibfnamefont {I.~K.}\ \bibnamefont {Bdikin}},
  \bibinfo {author} {\bibfnamefont {M.}~\bibnamefont {Kosec}}, \ and\ \bibinfo
  {author} {\bibfnamefont {A.~L.}\ \bibnamefont {Kholkin}},\ }\href
  {https://aip.scitation.org/doi/10.1063/1.3624810} {\bibfield  {journal}
  {\bibinfo  {journal} {J. Appl. Phys.}\ }\textbf {\bibinfo {volume} {110}},\
  \bibinfo {pages} {052001} (\bibinfo {year} {2011})}\BibitemShut {NoStop}%
\bibitem [{\citenamefont {Xiao}\ \emph {et~al.}(2013)\citenamefont {Xiao},
  \citenamefont {Poddar}, \citenamefont {Ducharme},\ and\ \citenamefont
  {Hong}}]{xiao_apl_13_PVDF_roughness_creep}%
  \BibitemOpen
  \bibfield  {author} {\bibinfo {author} {\bibfnamefont {Z.}~\bibnamefont
  {Xiao}}, \bibinfo {author} {\bibfnamefont {S.}~\bibnamefont {Poddar}},
  \bibinfo {author} {\bibfnamefont {S.}~\bibnamefont {Ducharme}}, \ and\
  \bibinfo {author} {\bibfnamefont {X.}~\bibnamefont {Hong}},\ }\href
  {https://doi.org/10.1063/1.4820784} {\bibfield  {journal} {\bibinfo
  {journal} {Appl. Phys. Lett.}\ }\textbf {\bibinfo {volume} {103}},\ \bibinfo
  {pages} {112903} (\bibinfo {year} {2013})}\BibitemShut {NoStop}%
\bibitem [{\citenamefont {Schmittbuhl}\ \emph
  {et~al.}(1995{\natexlab{a}})\citenamefont {Schmittbuhl}, \citenamefont
  {Vilotte},\ and\ \citenamefont
  {Roux}}]{schmittbuhl_pre_95_reliability_roughness}%
  \BibitemOpen
  \bibfield  {author} {\bibinfo {author} {\bibfnamefont {J.}~\bibnamefont
  {Schmittbuhl}}, \bibinfo {author} {\bibfnamefont {J.-P.}\ \bibnamefont
  {Vilotte}}, \ and\ \bibinfo {author} {\bibfnamefont {S.}~\bibnamefont
  {Roux}},\ }\href {http://link.aps.org/doi/10.1103/PhysRevE.51.131} {\bibfield
   {journal} {\bibinfo  {journal} {Phys. Rev. E}\ }\textbf {\bibinfo {volume}
  {51}},\ \bibinfo {pages} {131} (\bibinfo {year}
  {1995}{\natexlab{a}})}\BibitemShut {NoStop}%
\bibitem [{\citenamefont {Guyonnet}\ \emph {et~al.}(2012)\citenamefont
  {Guyonnet}, \citenamefont {Agoritsas}, \citenamefont {Bustingorry},
  \citenamefont {Giamarchi},\ and\ \citenamefont
  {Paruch}}]{guyonnet_prl_12_multiscaling}%
  \BibitemOpen
  \bibfield  {author} {\bibinfo {author} {\bibfnamefont {J.}~\bibnamefont
  {Guyonnet}}, \bibinfo {author} {\bibfnamefont {E.}~\bibnamefont {Agoritsas}},
  \bibinfo {author} {\bibfnamefont {S.}~\bibnamefont {Bustingorry}}, \bibinfo
  {author} {\bibfnamefont {T.}~\bibnamefont {Giamarchi}}, \ and\ \bibinfo
  {author} {\bibfnamefont {P.}~\bibnamefont {Paruch}},\ }\href
  {http://link.aps.org/doi/10.1103/PhysRevLett.109.147601} {\bibfield
  {journal} {\bibinfo  {journal} {Phys. Rev. Lett.}\ }\textbf {\bibinfo
  {volume} {109}},\ \bibinfo {pages} {147601} (\bibinfo {year}
  {2012})}\BibitemShut {NoStop}%
\bibitem [{\citenamefont {Rosso}\ \emph {et~al.}(2003)\citenamefont {Rosso},
  \citenamefont {Hartmann},\ and\ \citenamefont
  {Krauth}}]{rosso_prl_03_depinning}%
  \BibitemOpen
  \bibfield  {author} {\bibinfo {author} {\bibfnamefont {A.}~\bibnamefont
  {Rosso}}, \bibinfo {author} {\bibfnamefont {A.}~\bibnamefont {Hartmann}}, \
  and\ \bibinfo {author} {\bibfnamefont {W.}~\bibnamefont {Krauth}},\ }\href
  {https://link.aps.org/doi/10.1103/PhysRevE.67.021602} {\bibfield  {journal}
  {\bibinfo  {journal} {Phys. Rev. E}\ }\textbf {\bibinfo {volume} {67}},\
  \bibinfo {pages} {021602} (\bibinfo {year} {2003})}\BibitemShut {NoStop}%
\bibitem [{\citenamefont {Bustingorry}\ \emph {et~al.}(2012)\citenamefont
  {Bustingorry}, \citenamefont {Kolton},\ and\ \citenamefont
  {Giamarchi}}]{bustingorry_prb_12_depinning}%
  \BibitemOpen
  \bibfield  {author} {\bibinfo {author} {\bibfnamefont {S.}~\bibnamefont
  {Bustingorry}}, \bibinfo {author} {\bibfnamefont {A.}~\bibnamefont {Kolton}},
  \ and\ \bibinfo {author} {\bibfnamefont {T.}~\bibnamefont {Giamarchi}},\
  }\href {http://link.aps.org/doi/10.1103/PhysRevB.85.214416} {\bibfield
  {journal} {\bibinfo  {journal} {Phys. Rev. B}\ }\textbf {\bibinfo {volume}
  {85}},\ \bibinfo {pages} {214416} (\bibinfo {year} {2012})}\BibitemShut
  {NoStop}%
\bibitem [{\citenamefont {Leschhorn}\ and\ \citenamefont
  {Tang}(1993)}]{leschhorn_prl_93_superroughening}%
  \BibitemOpen
  \bibfield  {author} {\bibinfo {author} {\bibfnamefont {H.}~\bibnamefont
  {Leschhorn}}\ and\ \bibinfo {author} {\bibfnamefont {L.-H.}\ \bibnamefont
  {Tang}},\ }\href {https://link.aps.org/doi/10.1103/PhysRevLett.70.2973}
  {\bibfield  {journal} {\bibinfo  {journal} {Phys. Rev. Lett.}\ }\textbf
  {\bibinfo {volume} {70}},\ \bibinfo {pages} {2973} (\bibinfo {year}
  {1993})}\BibitemShut {NoStop}%
\bibitem [{\citenamefont {Shur}\ \emph {et~al.}(1991)\citenamefont {Shur},
  \citenamefont {Gruverman}, \citenamefont {Ponomarev}, \citenamefont
  {Rumyantsev},\ and\ \citenamefont
  {Tonkacheva}}]{shur_jetp_91_skyrmions_ferroelectric}%
  \BibitemOpen
  \bibfield  {author} {\bibinfo {author} {\bibfnamefont {V.~Y.}\ \bibnamefont
  {Shur}}, \bibinfo {author} {\bibfnamefont {A.~L.}\ \bibnamefont {Gruverman}},
  \bibinfo {author} {\bibfnamefont {N.~Y.}\ \bibnamefont {Ponomarev}}, \bibinfo
  {author} {\bibfnamefont {E.~L.}\ \bibnamefont {Rumyantsev}}, \ and\ \bibinfo
  {author} {\bibfnamefont {N.~A.}\ \bibnamefont {Tonkacheva}},\ }\href
  {http://www.jetpletters.ac.ru/ps/1169/article_17658.shtml} {\bibfield
  {journal} {\bibinfo  {journal} {JETP Lett.}\ }\textbf {\bibinfo {volume}
  {53}},\ \bibinfo {pages} {615} (\bibinfo {year} {1991})}\BibitemShut
  {NoStop}%
\bibitem [{\citenamefont {Dawber}\ \emph {et~al.}(2006)\citenamefont {Dawber},
  \citenamefont {Gruverman},\ and\ \citenamefont
  {Scott}}]{dawber_jpcm_06_DW_skyrmions}%
  \BibitemOpen
  \bibfield  {author} {\bibinfo {author} {\bibfnamefont {M.}~\bibnamefont
  {Dawber}}, \bibinfo {author} {\bibfnamefont {A.}~\bibnamefont {Gruverman}}, \
  and\ \bibinfo {author} {\bibfnamefont {J.~F.}\ \bibnamefont {Scott}},\ }\href
  {https://iopscience.iop.org/article/10.1088/0953-8984/18/5/L03} {\bibfield
  {journal} {\bibinfo  {journal} {J. Phys.: Condens. Matter}\ }\textbf
  {\bibinfo {volume} {18}},\ \bibinfo {pages} {L71} (\bibinfo {year}
  {2006})}\BibitemShut {NoStop}%
\bibitem [{\citenamefont {Brazovskii}\ and\ \citenamefont
  {Nattermann}(2004)}]{Brazovskii2004}%
  \BibitemOpen
  \bibfield  {author} {\bibinfo {author} {\bibfnamefont {S.}~\bibnamefont
  {Brazovskii}}\ and\ \bibinfo {author} {\bibfnamefont {T.}~\bibnamefont
  {Nattermann}},\ }\href {https://doi.org/10.1080/00018730410001684197}
  {\bibfield  {journal} {\bibinfo  {journal} {Adv. Phys.}\ }\textbf {\bibinfo
  {volume} {53}},\ \bibinfo {pages} {177} (\bibinfo {year} {2004})}\BibitemShut
  {NoStop}%
\bibitem [{\citenamefont {M{\aa}l{\o}y}\ \emph {et~al.}(1992)\citenamefont
  {M{\aa}l{\o}y}, \citenamefont {Hansen}, \citenamefont {Hinrichsen},\ and\
  \citenamefont {Roux}}]{maloy_prl_92_crack}%
  \BibitemOpen
  \bibfield  {author} {\bibinfo {author} {\bibfnamefont {K.~J.}\ \bibnamefont
  {M{\aa}l{\o}y}}, \bibinfo {author} {\bibfnamefont {A.}~\bibnamefont
  {Hansen}}, \bibinfo {author} {\bibfnamefont {E.~L.}\ \bibnamefont
  {Hinrichsen}}, \ and\ \bibinfo {author} {\bibfnamefont {S.}~\bibnamefont
  {Roux}},\ }\href {https://link.aps.org/doi/10.1103/PhysRevLett.68.213}
  {\bibfield  {journal} {\bibinfo  {journal} {Phys. Rev. Lett.}\ }\textbf
  {\bibinfo {volume} {68}},\ \bibinfo {pages} {213} (\bibinfo {year}
  {1992})}\BibitemShut {NoStop}%
\bibitem [{\citenamefont {Schmittbuhl}\ \emph
  {et~al.}(1995{\natexlab{b}})\citenamefont {Schmittbuhl}, \citenamefont
  {Schmitt},\ and\ \citenamefont
  {Scholz}}]{schmittbuhl_jgeophys_95_crack_scaling}%
  \BibitemOpen
  \bibfield  {author} {\bibinfo {author} {\bibfnamefont {J.}~\bibnamefont
  {Schmittbuhl}}, \bibinfo {author} {\bibfnamefont {F.}~\bibnamefont
  {Schmitt}}, \ and\ \bibinfo {author} {\bibfnamefont {C.}~\bibnamefont
  {Scholz}},\ }\href {https://doi.org/10.1029/94JB02885} {\bibfield  {journal}
  {\bibinfo  {journal} {J. Geophys. Res.}\ }\textbf {\bibinfo {volume} {100}},\
  \bibinfo {pages} {5953} (\bibinfo {year} {1995}{\natexlab{b}})}\BibitemShut
  {NoStop}%
\bibitem [{\citenamefont {Family}\ and\ \citenamefont
  {Vicsek}(1985)}]{FamilyVicsek}%
  \BibitemOpen
  \bibfield  {author} {\bibinfo {author} {\bibfnamefont {F.}~\bibnamefont
  {Family}}\ and\ \bibinfo {author} {\bibfnamefont {T.}~\bibnamefont
  {Vicsek}},\ }\href {http://iopscience.iop.org/0305-4470/18/2/005/} {\bibfield
   {journal} {\bibinfo  {journal} {J. Phys. A}\ }\textbf {\bibinfo {volume}
  {18}},\ \bibinfo {pages} {L75} (\bibinfo {year} {1985})}\BibitemShut
  {NoStop}%
\bibitem [{\citenamefont {L\'opez}\ \emph {et~al.}(1997)\citenamefont
  {L\'opez}, \citenamefont {Rodr\'{\i}guez},\ and\ \citenamefont
  {Cuerno}}]{lopez_pre_97_anomalous_scaling}%
  \BibitemOpen
  \bibfield  {author} {\bibinfo {author} {\bibfnamefont {J.~M.}\ \bibnamefont
  {L\'opez}}, \bibinfo {author} {\bibfnamefont {M.~A.}\ \bibnamefont
  {Rodr\'{\i}guez}}, \ and\ \bibinfo {author} {\bibfnamefont {R.}~\bibnamefont
  {Cuerno}},\ }\href {https://link.aps.org/doi/10.1103/PhysRevE.56.3993}
  {\bibfield  {journal} {\bibinfo  {journal} {Phys. Rev. E}\ }\textbf {\bibinfo
  {volume} {56}},\ \bibinfo {pages} {3993} (\bibinfo {year}
  {1997})}\BibitemShut {NoStop}%
\bibitem [{\citenamefont {L\'opez}\ and\ \citenamefont
  {Schmittbuhl}(1998)}]{lopez_pre_98_anomalous_scaling}%
  \BibitemOpen
  \bibfield  {author} {\bibinfo {author} {\bibfnamefont {J.~M.}\ \bibnamefont
  {L\'opez}}\ and\ \bibinfo {author} {\bibfnamefont {J.}~\bibnamefont
  {Schmittbuhl}},\ }\href {https://link.aps.org/doi/10.1103/PhysRevE.57.6405}
  {\bibfield  {journal} {\bibinfo  {journal} {Phys. Rev. E}\ }\textbf {\bibinfo
  {volume} {57}},\ \bibinfo {pages} {6405} (\bibinfo {year}
  {1998})}\BibitemShut {NoStop}%
\bibitem [{\citenamefont {Ramasco}\ \emph {et~al.}(2000)\citenamefont
  {Ramasco}, \citenamefont {L\'opez},\ and\ \citenamefont
  {Rodr\'iguez}}]{ramasco_prl_00_generic_scaling}%
  \BibitemOpen
  \bibfield  {author} {\bibinfo {author} {\bibfnamefont {J.~J.}\ \bibnamefont
  {Ramasco}}, \bibinfo {author} {\bibfnamefont {J.~M.}\ \bibnamefont
  {L\'opez}}, \ and\ \bibinfo {author} {\bibfnamefont {M.~A.}\ \bibnamefont
  {Rodr\'iguez}},\ }\href
  {https://link.aps.org/doi/10.1103/PhysRevLett.84.2199} {\bibfield  {journal}
  {\bibinfo  {journal} {Phys. Rev. Lett.}\ }\textbf {\bibinfo {volume} {84}},\
  \bibinfo {pages} {2199} (\bibinfo {year} {2000})}\BibitemShut {NoStop}%
\bibitem [{\citenamefont {Jost}\ \emph {et~al.}(1998)\citenamefont {Jost},
  \citenamefont {Heimel},\ and\ \citenamefont
  {Kleinefeld}}]{Jost_pre_98_roughness_B}%
  \BibitemOpen
  \bibfield  {author} {\bibinfo {author} {\bibfnamefont {M.}~\bibnamefont
  {Jost}}, \bibinfo {author} {\bibfnamefont {J.}~\bibnamefont {Heimel}}, \ and\
  \bibinfo {author} {\bibfnamefont {T.}~\bibnamefont {Kleinefeld}},\ }\href
  {\doibase 10.1103/PhysRevB.57.5316} {\bibfield  {journal} {\bibinfo
  {journal} {Phys. Rev. B}\ }\textbf {\bibinfo {volume} {57}},\ \bibinfo
  {pages} {5316} (\bibinfo {year} {1998})}\BibitemShut {NoStop}%
\bibitem [{\citenamefont {Paruch}\ \emph {et~al.}(2012)\citenamefont {Paruch},
  \citenamefont {Kolton}, \citenamefont {Hong}, \citenamefont {Ahn},\ and\
  \citenamefont {Giamarchi}}]{paruch_prb_12_quench}%
  \BibitemOpen
  \bibfield  {author} {\bibinfo {author} {\bibfnamefont {P.}~\bibnamefont
  {Paruch}}, \bibinfo {author} {\bibfnamefont {A.~B.}\ \bibnamefont {Kolton}},
  \bibinfo {author} {\bibfnamefont {X.}~\bibnamefont {Hong}}, \bibinfo {author}
  {\bibfnamefont {C.~H.}\ \bibnamefont {Ahn}}, \ and\ \bibinfo {author}
  {\bibfnamefont {T.}~\bibnamefont {Giamarchi}},\ }\href
  {http://link.aps.org/doi/10.1103/PhysRevB.85.214115} {\bibfield  {journal}
  {\bibinfo  {journal} {Phys. Rev. B}\ }\textbf {\bibinfo {volume} {85}},\
  \bibinfo {pages} {214115} (\bibinfo {year} {2012})}\BibitemShut {NoStop}%
\bibitem [{\citenamefont {Blaser}\ and\ \citenamefont
  {Paruch}(2012)}]{blaser_apl_12_CNT_FE}%
  \BibitemOpen
  \bibfield  {author} {\bibinfo {author} {\bibfnamefont {C.}~\bibnamefont
  {Blaser}}\ and\ \bibinfo {author} {\bibfnamefont {P.}~\bibnamefont
  {Paruch}},\ }\href {https://doi.org/10.1063/1.4757880} {\bibfield  {journal}
  {\bibinfo  {journal} {Appl. Phys. Lett.}\ }\textbf {\bibinfo {volume}
  {101}},\ \bibinfo {pages} {142906} (\bibinfo {year} {2012})}\BibitemShut
  {NoStop}%
\bibitem [{\citenamefont {Paruch}\ and\ \citenamefont
  {Triscone}(2006)}]{paruch_apl_06_stability}%
  \BibitemOpen
  \bibfield  {author} {\bibinfo {author} {\bibfnamefont {P.}~\bibnamefont
  {Paruch}}\ and\ \bibinfo {author} {\bibfnamefont {J.-M.}\ \bibnamefont
  {Triscone}},\ }\href {https://aip.scitation.org/doi/10.1063/1.2196482}
  {\bibfield  {journal} {\bibinfo  {journal} {Appl. Phys. Lett.}\ }\textbf
  {\bibinfo {volume} {88}},\ \bibinfo {pages} {162907} (\bibinfo {year}
  {2006})}\BibitemShut {NoStop}%
\bibitem [{\citenamefont {Kolton}\ \emph {et~al.}(2006)\citenamefont {Kolton},
  \citenamefont {Rosso}, \citenamefont {Giamarchi},\ and\ \citenamefont
  {Krauth}}]{kolton_prl_06_DWdepinning}%
  \BibitemOpen
  \bibfield  {author} {\bibinfo {author} {\bibfnamefont {A.~B.}\ \bibnamefont
  {Kolton}}, \bibinfo {author} {\bibfnamefont {A.}~\bibnamefont {Rosso}},
  \bibinfo {author} {\bibfnamefont {T.}~\bibnamefont {Giamarchi}}, \ and\
  \bibinfo {author} {\bibfnamefont {W.}~\bibnamefont {Krauth}},\ }\href
  {\doibase 10.1103/PhysRevLett.97.057001} {\bibfield  {journal} {\bibinfo
  {journal} {Phys. Rev. Lett.}\ }\textbf {\bibinfo {volume} {97}},\ \bibinfo
  {pages} {057001} (\bibinfo {year} {2006})}\BibitemShut {NoStop}%
\bibitem [{\citenamefont {Rosso}\ \emph {et~al.}(2007)\citenamefont {Rosso},
  \citenamefont {Le~Doussal},\ and\ \citenamefont
  {Wiese}}]{Rosso_prb_07_numericalFRG}%
  \BibitemOpen
  \bibfield  {author} {\bibinfo {author} {\bibfnamefont {A.}~\bibnamefont
  {Rosso}}, \bibinfo {author} {\bibfnamefont {P.}~\bibnamefont {Le~Doussal}}, \
  and\ \bibinfo {author} {\bibfnamefont {K.~J.}\ \bibnamefont {Wiese}},\ }\href
  {\doibase 10.1103/PhysRevB.75.220201} {\bibfield  {journal} {\bibinfo
  {journal} {Phys. Rev. B}\ }\textbf {\bibinfo {volume} {75}},\ \bibinfo
  {pages} {220201(R)} (\bibinfo {year} {2007})}\BibitemShut {NoStop}%
\bibitem [{\citenamefont {Ferrero}\ \emph
  {et~al.}(2013{\natexlab{a}})\citenamefont {Ferrero}, \citenamefont
  {Bustingorry}, \citenamefont {Kolton},\ and\ \citenamefont
  {Rosso}}]{Ferrero2013}%
  \BibitemOpen
  \bibfield  {author} {\bibinfo {author} {\bibfnamefont {E.~E.}\ \bibnamefont
  {Ferrero}}, \bibinfo {author} {\bibfnamefont {S.}~\bibnamefont
  {Bustingorry}}, \bibinfo {author} {\bibfnamefont {A.~B.}\ \bibnamefont
  {Kolton}}, \ and\ \bibinfo {author} {\bibfnamefont {A.}~\bibnamefont
  {Rosso}},\ }\href {http://dx.doi.org/10.1016/j.crhy.2013.08.002} {\bibfield
  {journal} {\bibinfo  {journal} {C. R. Physique}\ }\textbf {\bibinfo {volume}
  {14}},\ \bibinfo {pages} {641} (\bibinfo {year}
  {2013}{\natexlab{a}})}\BibitemShut {NoStop}%
\bibitem [{\citenamefont {Torres}\ and\ \citenamefont
  {Buceta}(2013)}]{Torres_epjb_13_anomalous_scaling}%
  \BibitemOpen
  \bibfield  {author} {\bibinfo {author} {\bibfnamefont {M.~F.}\ \bibnamefont
  {Torres}}\ and\ \bibinfo {author} {\bibfnamefont {R.~C.}\ \bibnamefont
  {Buceta}},\ }\href {\doibase 10.1140/epjb/e2012-30482-6} {\bibfield
  {journal} {\bibinfo  {journal} {Eur. Phys. J. B}\ }\textbf {\bibinfo {volume}
  {86}},\ \bibinfo {pages} {20} (\bibinfo {year} {2013})}\BibitemShut {NoStop}%
\bibitem [{\citenamefont {Ferrero}\ \emph
  {et~al.}(2013{\natexlab{b}})\citenamefont {Ferrero}, \citenamefont
  {Bustingorry},\ and\ \citenamefont
  {Kolton}}]{Ferrero_pre_13_numerical_exponents}%
  \BibitemOpen
  \bibfield  {author} {\bibinfo {author} {\bibfnamefont {E.~E.}\ \bibnamefont
  {Ferrero}}, \bibinfo {author} {\bibfnamefont {S.}~\bibnamefont
  {Bustingorry}}, \ and\ \bibinfo {author} {\bibfnamefont {A.~B.}\ \bibnamefont
  {Kolton}},\ }\href {\doibase 10.1103/PhysRevE.87.032122} {\bibfield
  {journal} {\bibinfo  {journal} {Phys. Rev. E}\ }\textbf {\bibinfo {volume}
  {87}},\ \bibinfo {pages} {032122} (\bibinfo {year}
  {2013}{\natexlab{b}})}\BibitemShut {NoStop}%
\bibitem [{\citenamefont {Grassi}\ \emph {et~al.}(2018)\citenamefont {Grassi},
  \citenamefont {Kolton}, \citenamefont {Jeudy}, \citenamefont {Mougin},
  \citenamefont {Bustingorry},\ and\ \citenamefont {Curiale}}]{Grassi2018}%
  \BibitemOpen
  \bibfield  {author} {\bibinfo {author} {\bibfnamefont {M.~P.}\ \bibnamefont
  {Grassi}}, \bibinfo {author} {\bibfnamefont {A.~B.}\ \bibnamefont {Kolton}},
  \bibinfo {author} {\bibfnamefont {V.}~\bibnamefont {Jeudy}}, \bibinfo
  {author} {\bibfnamefont {A.}~\bibnamefont {Mougin}}, \bibinfo {author}
  {\bibfnamefont {S.}~\bibnamefont {Bustingorry}}, \ and\ \bibinfo {author}
  {\bibfnamefont {J.}~\bibnamefont {Curiale}},\ }\href
  {https://link.aps.org/doi/10.1103/PhysRevB.98.224201} {\bibfield  {journal}
  {\bibinfo  {journal} {Phys. Rev. B}\ }\textbf {\bibinfo {volume} {98}},\
  \bibinfo {pages} {224201} (\bibinfo {year} {2018})}\BibitemShut {NoStop}%
\bibitem [{\citenamefont {Agoritsas}\ \emph {et~al.}(2010)\citenamefont
  {Agoritsas}, \citenamefont {Lecomte},\ and\ \citenamefont
  {Giamarchi}}]{agoritsas_2010_PhysRevB_82_184207}%
  \BibitemOpen
  \bibfield  {author} {\bibinfo {author} {\bibfnamefont {E.}~\bibnamefont
  {Agoritsas}}, \bibinfo {author} {\bibfnamefont {V.}~\bibnamefont {Lecomte}},
  \ and\ \bibinfo {author} {\bibfnamefont {T.}~\bibnamefont {Giamarchi}},\
  }\href {\doibase 10.1103/PhysRevB.82.184207} {\bibfield  {journal} {\bibinfo
  {journal} {Phys. Rev. B}\ }\textbf {\bibinfo {volume} {82}},\ \bibinfo
  {pages} {184207} (\bibinfo {year} {2010})}\BibitemShut {NoStop}%
\bibitem [{\citenamefont {Agoritsas}\ \emph {et~al.}(2013)\citenamefont
  {Agoritsas}, \citenamefont {Lecomte},\ and\ \citenamefont
  {Giamarchi}}]{agoritsas_2012_FHHtri-numerics}%
  \BibitemOpen
  \bibfield  {author} {\bibinfo {author} {\bibfnamefont {E.}~\bibnamefont
  {Agoritsas}}, \bibinfo {author} {\bibfnamefont {V.}~\bibnamefont {Lecomte}},
  \ and\ \bibinfo {author} {\bibfnamefont {T.}~\bibnamefont {Giamarchi}},\
  }\href {\doibase 10.1103/PhysRevE.87.062405} {\bibfield  {journal} {\bibinfo
  {journal} {Phys. Rev. E}\ }\textbf {\bibinfo {volume} {87}},\ \bibinfo
  {pages} {062405} (\bibinfo {year} {2013})}\BibitemShut {NoStop}%
\bibitem [{\citenamefont {Huse}\ \emph {et~al.}(1985)\citenamefont {Huse},
  \citenamefont {Henley},\ and\ \citenamefont
  {Fisher}}]{huse_henley_fisher_1985_PhysRevLett55_2924}%
  \BibitemOpen
  \bibfield  {author} {\bibinfo {author} {\bibfnamefont {D.~A.}\ \bibnamefont
  {Huse}}, \bibinfo {author} {\bibfnamefont {C.~L.}\ \bibnamefont {Henley}}, \
  and\ \bibinfo {author} {\bibfnamefont {D.~S.}\ \bibnamefont {Fisher}},\
  }\href {\doibase 10.1103/PhysRevLett.55.2924} {\bibfield  {journal} {\bibinfo
   {journal} {Phys. Rev. Lett.}\ }\textbf {\bibinfo {volume} {55}},\ \bibinfo
  {pages} {2924} (\bibinfo {year} {1985})}\BibitemShut {NoStop}%
\bibitem [{\citenamefont {M\'ezard}\ and\ \citenamefont
  {Parisi}(1992)}]{mezard_jdpi_92_manifolds}%
  \BibitemOpen
  \bibfield  {author} {\bibinfo {author} {\bibfnamefont {M.}~\bibnamefont
  {M\'ezard}}\ and\ \bibinfo {author} {\bibfnamefont {G.}~\bibnamefont
  {Parisi}},\ }\href {https://doi.org/10.1051/jp1:1992278} {\bibfield
  {journal} {\bibinfo  {journal} {J. de Phys. I}\ }\textbf {\bibinfo {volume}
  {2}},\ \bibinfo {pages} {2231} (\bibinfo {year} {1992})}\BibitemShut
  {NoStop}%
\bibitem [{\citenamefont {Kardar}(1985)}]{Kardar_prl_85_DP}%
  \BibitemOpen
  \bibfield  {author} {\bibinfo {author} {\bibfnamefont {M.}~\bibnamefont
  {Kardar}},\ }\href {\doibase 10.1103/PhysRevLett.55.2923} {\bibfield
  {journal} {\bibinfo  {journal} {Phys. Rev. Lett.}\ }\textbf {\bibinfo
  {volume} {55}},\ \bibinfo {pages} {2923} (\bibinfo {year}
  {1985})}\BibitemShut {NoStop}%
\bibitem [{\citenamefont {Santucci}\ \emph {et~al.}(2007)\citenamefont
  {Santucci}, \citenamefont {al\o y}, \citenamefont {Delaplace}, \citenamefont
  {Mathiesen}, \citenamefont {Hansen}, \citenamefont {Bakke}, \citenamefont
  {Schmittbuhl}, \citenamefont {Vanel},\ and\ \citenamefont
  {Purusattam}}]{santucci_pre_07_fracture_statistics}%
  \BibitemOpen
  \bibfield  {author} {\bibinfo {author} {\bibfnamefont {S.}~\bibnamefont
  {Santucci}}, \bibinfo {author} {\bibfnamefont {K.~J.~M.}\ \bibnamefont {al\o
  y}}, \bibinfo {author} {\bibfnamefont {A.}~\bibnamefont {Delaplace}},
  \bibinfo {author} {\bibfnamefont {J.}~\bibnamefont {Mathiesen}}, \bibinfo
  {author} {\bibfnamefont {A.}~\bibnamefont {Hansen}}, \bibinfo {author}
  {\bibfnamefont {J.~O.~H.}\ \bibnamefont {Bakke}}, \bibinfo {author}
  {\bibfnamefont {J.}~\bibnamefont {Schmittbuhl}}, \bibinfo {author}
  {\bibfnamefont {L.}~\bibnamefont {Vanel}}, \ and\ \bibinfo {author}
  {\bibfnamefont {R.}~\bibnamefont {Purusattam}},\ }\href
  {http://link.aps.org/doi/10.1103/PhysRevE.75.016104} {\bibfield  {journal}
  {\bibinfo  {journal} {Phys. Rev. E}\ }\textbf {\bibinfo {volume} {75}},\
  \bibinfo {pages} {016104} (\bibinfo {year} {2007})}\BibitemShut {NoStop}%
\bibitem [{\citenamefont {Middleton}(1992)}]{Middleton1992}%
  \BibitemOpen
  \bibfield  {author} {\bibinfo {author} {\bibfnamefont {A.~A.}\ \bibnamefont
  {Middleton}},\ }\href {\doibase 10.1103/PhysRevLett.68.670} {\bibfield
  {journal} {\bibinfo  {journal} {Phys. Rev. Lett.}\ }\textbf {\bibinfo
  {volume} {68}},\ \bibinfo {pages} {670} (\bibinfo {year} {1992})}\BibitemShut
  {NoStop}%
\bibitem [{\citenamefont {Bustingorry}\ \emph {et~al.}(2010)\citenamefont
  {Bustingorry}, \citenamefont {Kolton},\ and\ \citenamefont
  {Giamarchi}}]{Bustingorry_prb_10_random_periodic}%
  \BibitemOpen
  \bibfield  {author} {\bibinfo {author} {\bibfnamefont {S.}~\bibnamefont
  {Bustingorry}}, \bibinfo {author} {\bibfnamefont {A.~B.}\ \bibnamefont
  {Kolton}}, \ and\ \bibinfo {author} {\bibfnamefont {T.}~\bibnamefont
  {Giamarchi}},\ }\href {\doibase 10.1103/PhysRevB.82.094202} {\bibfield
  {journal} {\bibinfo  {journal} {Phys. Rev. B}\ }\textbf {\bibinfo {volume}
  {82}},\ \bibinfo {pages} {094202} (\bibinfo {year} {2010})}\BibitemShut
  {NoStop}%
\bibitem [{\citenamefont {Bustingorry}\ and\ \citenamefont
  {Kolton}(2010)}]{Bustingorry_pip_10_random_periodic}%
  \BibitemOpen
  \bibfield  {author} {\bibinfo {author} {\bibfnamefont {S.}~\bibnamefont
  {Bustingorry}}\ and\ \bibinfo {author} {\bibfnamefont {A.}~\bibnamefont
  {Kolton}},\ }\href
  {http://www.papersinphysics.org/papersinphysics/article/view/44} {\bibfield
  {journal} {\bibinfo  {journal} {Papers in Physics}\ }\textbf {\bibinfo
  {volume} {2}} (\bibinfo {year} {2010})}\BibitemShut {NoStop}%
\bibitem [{\citenamefont {Kolton}\ \emph {et~al.}(2013)\citenamefont {Kolton},
  \citenamefont {Bustingorry}, \citenamefont {Ferrero},\ and\ \citenamefont
  {Rosso}}]{Kolton_jstat_13_fc_random_periodic}%
  \BibitemOpen
  \bibfield  {author} {\bibinfo {author} {\bibfnamefont {A.~B.}\ \bibnamefont
  {Kolton}}, \bibinfo {author} {\bibfnamefont {S.}~\bibnamefont {Bustingorry}},
  \bibinfo {author} {\bibfnamefont {E.~E.}\ \bibnamefont {Ferrero}}, \ and\
  \bibinfo {author} {\bibfnamefont {A.}~\bibnamefont {Rosso}},\ }\href
  {http://stacks.iop.org/1742-5468/2013/i=12/a=P12004} {\bibfield  {journal}
  {\bibinfo  {journal} {J. Stat. Mech.: Theor. Exp.}\ }\textbf {\bibinfo
  {volume} {2013}},\ \bibinfo {pages} {P12004} (\bibinfo {year}
  {2013})}\BibitemShut {NoStop}%
\end{thebibliography}

%

\end{document}